\documentclass[prc,twocolumn,showpacs,amsmath,amssymb,floatfix,superscriptaddress]{revtex4-1}
\usepackage{graphicx,graphics}
\usepackage{amsmath}
\usepackage{amssymb}
\usepackage{epsfig}
\usepackage{epstopdf}
\usepackage{amstext}
\usepackage{amsthm}
\usepackage{amsfonts}
\usepackage{latexsym}
\usepackage{array}
\usepackage{xfrac}
\usepackage{color}
\usepackage{subfigure}
\usepackage{multirow}
\usepackage{fontenc}
\usepackage{bm}
\usepackage{adjustbox}
\usepackage{dcolumn}
\usepackage{ulem}
\begin{document}

\title{Woods-Saxon-type of mean-field potentials with effective mass derived from the D1S Gogny force}

\author{A. Bhagwat}
\affiliation{School of Physical Sciences, UM-DAE Centre for Excellence in Basic Sciences, Mumbai 400 098, India}
\author{M. Centelles}
\affiliation{Departament de F\'isica Qu\`antica i Astrof\'isica and Institut de Ci\`encies del Cosmos, 
Facultat de F\'isica, Universitat de Barcelona,  Mart{\'i} i Franqu{\`e}s 1, E-08028 Barcelona, Spain}
\author{X. Vi\~nas}
\affiliation{Departament de F\'isica Qu\`antica i Astrof\'isica and Institut de Ci\`encies del Cosmos, 
Facultat de F\'isica, Universitat de Barcelona, Mart{\'i} i Franqu{\`e}s 1, E-08028 Barcelona, Spain}
\author{P. Schuck}
\affiliation{Institut de Physique Nucl\'eaire, IN2P3-CNRS, Universit\'e Paris-Sud,
F-91406 Orsay-C\'edex, France\\
Laboratoire de Physique et Mod\'elisation des Milieux Condens\'es, CNRS and 
Universit\'e Joseph Fourier, 25 Avenue des Martyrs,\\
Boite Postale 166, F-38042 Grenoble Cedex 9, France}

\date{\today}

\begin{abstract}
Analytic average mean-field potentials of the Fermi-function (Woods-Saxon) type for the whole nuclear chart 
with space-dependent effective mass are deduced from the D1S Gogny force. Those ready-for-use potentials are 
advertised as an alternative to other existing phenomenological mean-field potentials. 

\end{abstract}

\maketitle
\section{Introduction}

Phenomenological mean-field potentials of the Woods-Saxon type have been and are still in use for 
various purposes as, e.g., the low-cost evaluation
of shell effects in the microscopic-macroscopic 
(Mic-Mac) calculation of the nuclear mass table. The latter approach counts still among the most 
accurate ones for the prediction of nuclear masses. However, those phenomenological mean-field potentials generally 
are given without effective mass, i.e., they have the bare mass $m$ as input. It is, however, well-known that 
the real part of the optical potential is energy-dependent and, therefore, an important physical ingredient is 
missing in present-day phenomenological mean field potentials.  

This work intends to contribute to fill this gap. To this end we develop a phenomenological mean-field including
effective mass, which to our knowledge does not exist in the previous literature.
We deduce with semiclassical methods an analytic mean-field potential from the 
D1S Gogny force with space-dependent effective mass $m^*(r)$. 
Let us point out at this stage that the choice of the D1S Gogny force is not essential.
We could have taken as well a Skyrme force with effective mass. The choice of the
Gogny force is rather dictated because in the following paper we want to study its behavior
in conjunction with the set up of a modified Mic-Mac model.
Nevertheless, a possible advantage of using Gogny forces instead of the Skyrme ones is that in dynamical calculations, 
for example response function, one can treat higher momentum transfer whereas in the Skyrme case one is limited to 
low momenta, since it represents a quadratic expansion in momentum. Another possible advantage of using Gogny interactions 
may be the fact that these forces are designed such that they can describe simultaneously the particle-hole and the 
particle-particle channels, which allows one to describe succesfully open shell nuclei with the same interaction
wihtout incresing the number of adjustable parameters. 
This is not the case of the Skyrme interactions, where the pairing field is computed with a different force.
With our existing technology \cite{centelles98,gridnev98,soubbotin00}
of solving Extended Thomas-Fermi (ETF) integro-differential equations with density functionals up to second-order 
$\hbar$ corrections using a given (finite-range) effective force,
quantities such as mean-field potentials or effective masses can
be readily computed.
These potentials are subsequently accurately fitted by functions of the 
Fermi-type yielding analytic expressions ready for use. Since our approach has never been published in a 
coherent way, we will present in section II, the semiclassical theory which allows us to determine such mean-field 
potentials with effective mass. However, the reader not interested in the details of our approach, may directly 
read section III where we give the explicit analytical form of our potentials and effective masses. 
In a follow-up article, we will indeed show that this strategy can yield very accurate masses in a 
Mic-Mac calculation.

The paper is organized as follows. In the second section we review the ETF method with finite-range
interactions and apply this formalism to the specific case of Gogny forces. 
The third section is devoted to the 
fits to generalized Fermi distributions of the results of the ETF calculations, i.e., neutron and proton
densities, central and spin-orbit single-particle potentials and effective masses. These fits provide a simple 
parametrization of the aforementioned quantities as functions of the mass ($A$) and atomic ($Z$) numbers. These 
parametrizations may be useful for applications where parametrized densities, single-particle potentials or 
effective masses are required. 
In the fourth section we discuss some results for energy calculations.
Finally, our conclusions are given in the last section.

\section{The Extended Thomas-Fermi approximation with finite-range forces}
  
The ETF approximation to the Hartree-Fock (HF) method for non-local forces was introduced and 
discussed in Refs.\cite{centelles98,gridnev98,soubbotin00}. This approximation is rather 
general and can be applied to any finite-range force. However, in this work we will apply this
formalism to the particular case of the Gogny interaction \cite{decharge80}. 
Due to the general applicability
of our approximation, and given that there are several parametrizations of the Gogny interaction 
available in the literature, we do not specify the precise force here. Only later we will 
specialize to the D1S force \cite{berger91}. 
In general, the Gogny-type of force \cite{decharge80} can be
written as
\begin{eqnarray}
V({\bf R},{\bf s})= \sum_{i=1}^{i=2}
\left( W_i + B_iP^{\sigma} - H_iP^{\tau} - M_iP^{\sigma}P^{\tau}\right) v_i(s) 
\nonumber \\
+ \left[t_3\left(1 + x_3P^{\sigma}\right)\rho^{\alpha}({\bf R})
+ i W_0\left(\hat{\sigma}_1 + \hat{\sigma}_2\right)\cdot{\hat{\bf k}^\dagger} \times{\hat {\bf k}}\right]
\delta\left({\bf s}\right)\, , \nonumber \\
\label{eq1}
\end{eqnarray}
where $W_i$, $B_i$, $H_i$ and $M_i$ are the usual spin-isospin exchange strength parameters of the 
finite-range force and $v_i(s)=\exp(-s^2/\mu_i^2)$ (i=1,2) are the Gaussian form 
factors. The second term in Eq.~(\ref{eq1}) is the zero-range density-dependent contribution
and the last one corresponds to the spin-orbit interaction, which is also 
zero-range with a strength $W_0$ as in the case of Skyrme forces \cite{vautherin72}.  
In Eq.(\ref{eq1}) ${\bf R}=({\bf r_1}+{\bf r_2})/2$ and ${\bf s}= {\bf r_1}-{\bf r_2}$ are the 
center of mass and relative coordinates.
In the spin-orbit term the quantity ${\hat {\bf k}}$ represents the relative momentum of the two 
nucleons, expressed as ${\hat{\bf k}} = (\nabla_1 - \nabla_2)/2i$. 

The HF energy is calculated as the integral of the energy density ${\cal H}$, which can 
be split into its kinetic, nuclear, Coulomb and spin-orbit contributions. For a finite-range 
nuclear interaction we can write
\begin{widetext}
\begin{equation}
E_{HF} =  \int {\cal H} \, d{\bf R} = \int \left[{\cal H}_{kin} + {\cal H}^{nucl}_{dir} + {\cal H}^{nucl}_{exch}
+ {\cal H}^{nucl}_{z.range} + {\cal H}_{Coul} + {\cal H}_{s.o}\right] \, d{\bf R} \quad .
\label{eq2}
\end{equation}
\end{widetext}
 
The key quantity for writing the different contributions to the energy density is the one-body 
density matrix, which, at HF level, is defined as 
\begin{equation}
\rho \left({\bf R} + \frac{\bf s}{2},{\bf R} - \frac{\bf s}{2}\right)= 
\sum_{i=1}^{A} \phi_i^*\left({\bf r}\right)\phi_i \left({\bf r'}\right) \, ,
\label{eq3}
\end{equation}
where $\phi_i \left({\bf r}\right)$ are the single-particle HF wavefunctions. From this density matrix
the particle, kinetic energy and spin densities can be obtained as
\begin{eqnarray}
\rho\left({\bf R}\right) = \left. \rho\left({\bf R},{\bf s}\right)\right\vert_{s=0}
\label{eq4}
\end{eqnarray}
\begin{equation}
\tau({\bf R})= \left. \left(\frac{1}{4}\Delta_R - \Delta_s\right) \rho\left({\bf R},{\bf s}\right)\right\vert_{s=0} \, , 
\label{eq5}
\end{equation}
and 
\begin{equation}
{\bf J}({\bf R})= \left. -i\left[{\bf \hat{\sigma}} \times \left(\frac{1}{2}{\bf \nabla_R} + 
{\bf \nabla_s}\right)\right]\rho\left({\bf R},{\bf s}\right)\right \vert_{s=0},
\label{eq6}
\end{equation}
respectively.

The ETF approximation to the Hartree-Fock (HF) energy for non-local potentials consists of 
replacing the quantal HF  density matrix by its semiclassical counterpart. The latter contains, in 
addition to the pure Thomas-Fermi (TF)($\hbar^0$) term, the corrective gradient contributions 
up to the order $\hbar^2$ and even beyond, if so desired. Here we will consider the corrective 
terms only up to the $\hbar^2$ order.
In this approximation the ETF density matrix for each kind of particles,
which was derived in Refs.\cite{centelles98,gridnev98,soubbotin00}, 
can be written as  

\begin{widetext}
\begin{eqnarray}
&&{\tilde \rho}\left({\bf R},s\right) = \frac{3j_1\left(k_F s\right)}{k_F s}\rho 
+ \frac{s^2}{216}\Bigg[\left\{\left[\left(9 - 2k_F\frac{f_k}{f}
- 2k_F^2\frac{f_{kk}}{f} +  k_F^2\frac{f_k^2}{f}\right)\frac{j_1\left(k_F s\right)}{k_F s} -
4 j_0\left(k_F s\right)\right] 
\frac{\left({\bf \nabla}\rho\right)^2}{\rho} \, - \right. \nonumber \\
&& \left. \left[\left(18 +  6k_F\frac{f_k}{f}\right) \frac{j_1(k_F s)}{k_F s} - 3j_0(k_F s)\right]\Delta \rho
- \left[18\rho \frac{\Delta f}{f} + \left(18 - 6k_F \frac{f_k}{f}\right)
\frac{{\bf \nabla}\rho \cdot {\bf \nabla} f}{f}
+ 12 k_F \frac{{\bf \nabla}\rho \cdot {\bf \nabla} f_k}{f} - 9\rho \frac{\left({\bf \nabla} f\right)^2}{f} \right]
\frac{j_1(k_F s)}{k_F s}\right\}\nonumber \\
&&- \frac{m^2}{\hbar^4}\frac{\rho W^2}{f^2}s^2 \frac{j_1(k_F s)}{k_F s} -
\frac{im}{2\hbar^2}\frac{\rho}{f}\hat{\sigma}\cdot\left({\bf W}\times{\bf s}\right)
\frac{3j_1(k_F s)}{k_F s}\Bigg]_{k=k_F} \, .
\label{eq7}
\end{eqnarray}
\end{widetext}

Notice that for the sake of brevity, the subscripts `$q$' denoting particle type (neutron 
or proton) have been suppressed. Here, $k_F=\left(3 \pi^2 \rho({\bf R})\right)^{1/3}$ is the local Fermi 
momentum for each type of nucleons and $j_0(k_F s)$ and $j_1(k_F s)$ are the spherical Bessel functions of 
orders 0 and 1 respectively. The quantity $f=f({\bf R},k)$ in Eq.~(\ref{eq7})
is the inverse of the position and momentum-dependent effective mass, which at ETF-$\hbar^2$ level is defined as: 
\begin{equation}
f_q\left({\bf R},k\right) = 1 + \frac{m}{\hbar^2k}\frac{\partial V^{nucl}_{exch,q,0}({\bf R},k)}{\partial k},
\label{eq18}
\end{equation}
where $V^{nucl}_{exch,q,0}({\bf R},k)$ is the Wigner transform of the exchange potential at
TF level, which is given by
\begin{widetext}
\begin{equation}
V^{nucl}_{exch,q,0}({\bf R},k) = \sum_{i=1}^{i=2} \int \, d{\bf s} \, e^{-i{\bf k}\cdot{\bf s}}v_i(s)
\left[X_{e1,i}\rho_q({\bf R})\frac{3j_1(k_{F_q}({\bf R})s)}{k_{F_q}({\bf R})s}  
+ X_{e2,i}\rho_{q'}({\bf R})\frac{3j_1(k_{F_{q'}}({\bf R})s)}{k_{F_{q'}}({\bf R})s}\right] \, , 
\label{eq19}
\end{equation}
\end{widetext} 
where $q=n,p$ and $q'=p,n$, respectively. In this equation the coefficients 
$X_{e1,i}=M_i+H_i/2-B_i-W_i/2$ and $X_{e2,i}=M_i+H_i/2$ are combinations of the strengths of
the different exchange terms in Eq.(\ref{eq1}). 
It is important to note that the function
$f=f({\bf R},k)$ in Eq.~(\ref{eq7}) and its spatial derivatives $\nabla f \equiv \nabla_R f({\bf R},k)$
and $\Delta f \equiv \Delta_R f({\bf R},k)$, momentum derivatives $f_{k} \equiv f_{k}({\bf R},k)$
and $f_{kk}\equiv f_{kk}({\bf R},k)$ that represent the first and second
derivatives of $f({\bf R},k)$ with respect to $k$, and mixed derivatives  
$\nabla f_{k}\equiv \nabla_R f_{k}({\bf R},k)$ are all them  evaluated at $k=k_F$ after taking the
corresponding derivative. This is indicated by the subscript in the rightmost bracket of Eq.(\ref{eq7}).   
The last two terms in the density matrix (Eq. (\ref{eq7})) are the contributions due
to the spin-orbit force, which can be obtained from the propagator including this interaction 
(see e.g. Ref.\cite{bartel82}). These terms depend on the form-factor of the spin-orbit 
potential, which for each type of nucleons is given by:
\begin{equation}
{\bf W}_q = \frac{1}{2}W_0\big({\bf \nabla}\rho + {\bf \nabla}\rho_q\big).
\label{eq10}
\end{equation}

This ETF formalism is completely general for interactions with a finite-range central term, 
as for example the Gogny force given in Eq.(\ref{eq1}). The impact of the finite-range part
of the interaction on the density-matrix appears through the momentum and position-dependent effective 
mass defined in Eq.(\ref{eq18}) and its momentum and spatial derivatives. More details about 
the derivation of the ETF density matrix (Eq. (\ref{eq7})) and its comparison with earlier density 
matrix expansions available in the literature \cite{negele72,campi78} can be found in 
Ref.\cite{soubbotin00}.

The explicit form of the semiclassical kinetic energy and spin densities at ETF level can now
 easily be derived from the density matrix (Eq. (\ref{eq7})) using Eqs.(\ref{eq5}) and (\ref{eq6}),
respectively. The kinetic energy density is given by

\begin{eqnarray}
\tau \left({\bf R}\right) &=& \left. \left(\frac{1}{4}\Delta_R - \Delta_s \right) 
{\tilde \rho}\left({\bf R},s\right)\right\vert_{s=0}  \nonumber \\ 
 &=& \frac{3}{5} k_F^2 \rho 
+ \frac{1}{12}\Delta \rho \left[4 + \frac{2}{3}k_F \frac{f_k}{f} \right] \nonumber \\
&+&\frac{1}{36}\frac{\left({\bf \nabla} \rho\right)^2}{\rho} \left[ 1 + \frac{2}{3}k_F \frac{f_k}{f} +
\frac{2}{3}k_F^2 \frac{f_{kk}}{f}- \frac{1}{3}k_F^2 \frac{f_k^2}{f^2} \right]
\nonumber \\
&+&\frac{1}{6}\frac{\rho}{f} \left[\Delta f - \frac{({\bf \nabla}f)^2}{2f} \right] 
+ \frac{1}{6}\frac{{\bf \nabla}\rho \cdot {\bf \nabla}f}{f}
\left[1 - \frac{1}{3}k_F \frac{f_k}{f}\right] \nonumber \\
&+& \frac{1}{9}\frac{{\bf \nabla}\rho \cdot {\bf \nabla}f_k}{f}
+ \frac{1}{2}\left(\frac{2m}{\hbar^2}\right)^2 \frac{\rho}{f^2} W^2,
\label{eq8}
\end{eqnarray}
and the spin density at ETF level can be expressed as
\begin{widetext}
\begin{eqnarray}
{\bf J}_{ETF} &=& -i \left. \mathrm{Tr} \left\{ \left[\sigma \times \left(\frac{{\bf \nabla}_R}{2} + {\bf \nabla}_s)\right)\right]
\left(\frac{-im}{2\hbar^2}\right)\frac{\rho}{f} \left[\sigma \cdot ({\bf W}\times{\bf s})\right]\right\}
\frac{3j_1(k_F s)}{k_F s}\right\vert_{s=0}  \nonumber \\
&=& \left. -\frac{3m}{\hbar^2}\frac{\rho}{f}\left[{\bf \nabla_s} \times \left({\bf W}\times{\bf s}\right)\right]
\frac{3j_1(k_F s)}{k_F s}\right\vert_{s=0} = - \frac{2m}{\hbar^2}\frac{\rho {\bf W}}{f}.
\label{eq9}
\end{eqnarray}
\end{widetext}

Therefore, we see that at ETF level both, the kinetic energy and spin densities, become functionals of the 
particle densities $\rho_q({\bf R})$ for neutrons and protons. It is worthwhile to note that 
in the kinetic energy density (Eq. (\ref{eq8})) finite-range effects are encoded through the inverse effective 
mass $f$
(Eq. (\ref{eq18})) and its derivatives with respect to the position and momentum. If the effective mass
 is independent of the momentum, as it is the case of the zero-range Skyrme forces 
\cite{vautherin72}, the kinetic 
energy reduces to the $\hbar^2$ expansion reported in \cite{brack85} whereas, if the effective mass
 is equal to the physical one, the kinetic energy density reduces to the well-known 
Weizs\"acker term. The semiclassical kinetic energy and spin densities are used to
calculate the different pieces entering in the energy density (Eq. (\ref{eq2})), which for Gogny
interactions are reported in the Appendix 1.
It is important to note that due to the structure of the ETF density matrix (Eq. (\ref{eq7})), the 
exchange energy can be written as the sum of a $\hbar^0$ (Slater) contribution, which corresponds to 
the exchange energy in infinite nuclear matter, plus a $\hbar^2$ contribution, which can be finally 
recast in terms of the neutron and proton densities and their gradients up to second order, as it is 
explained in detail in Appendix 1. Therefore, in the ETF approximation 
\cite{soubbotin00}, the exchange energy becomes a local functional of the particle densities, 
as also happens starting from the Negele-Vautherin or the Campi-Bouyssy expansions of the density matrix 
\cite{negele72,campi78}. 
 
Taking into account the semiclassical kinetic energy and spin densities, the latter one used to
compute the spin-orbit energy, the total HF energy (Eq. (\ref{eq2})) at ETF level, $E_{HF}^{ETF}$, 
can be finally recast as a functional of the neutron and proton densities only (see Appendix 1 for 
further details). Therefore to find the semiclassical energy and the density profiles of a 
nucleus, one shall solve the following set of coupled equations of motion for neutrons and protons:
\begin{equation}
\frac{\delta}{\delta \rho_n} \bigg[ E_{HF}^{ETF} - \mu_n \int \rho_n({\bf R})
d{\bf R} \bigg] = 0 
\label{eq20}
\end{equation}
and
\begin{equation}
\frac{\delta}{\delta \rho_p} \bigg[ E_{HF}^{ETF} - \mu_p \int \rho_p({\bf R})
d{\bf R} \bigg] = 0. 
\label{eq21}
\end{equation}

This set of coupled second-order non-linear integro-differential equations can be solved 
for the neutron and proton densities by using, for example, the imaginary time-step method 
(see for example \cite{centelles90} and references therein). The self-consistent solution of  
Eqs.(\ref{eq20}) and (\ref{eq21}) provides the semiclassical proton and neutron densities, 
$\rho_n$ and $\rho_p$ respectively, which correspond to the fully variational solution of 
the ETF HF energy. An application of this semiclassical ETF method using the D1 Gogny force 
was reported in Ref.\cite{gridnev98}.

However, in this work instead of solving the full variational equations (\ref{eq20}) and 
(\ref{eq21}), which give the proton and neutron densities numerically point by point, we 
perform a restricted variational minimization of the ETF HF energy (Eq. (\ref{eq2})) 
using trial neutron and proton densities of generalized Fermi type for a large set of 
nuclei along the whole periodic table. In a second step, we fit the parameters of the 
trial densities, i.e. radii and diffuseness, as functions of the atomic mass and  
charge numbers. The semiclassical ETF approximation to obtain density profiles and binding 
energies in a restricted variational method with trial densities has been employed quite often 
in the past together with Skyrme forces \cite{brack85}. A detailed comparison between the 
predictions of the ETF approach using trial densities and solving the full equations of 
motion in the case of Skyrme forces can be found in Ref.\cite{centelles90}. From this study 
the following conclusions are obtained. On the one hand, it turned out that the density 
profiles using trial densities nicely reproduce the fully variational ones.
On the other hand, the energies derived through the restricted variational procedure lie
very close to energies computed from the fully variational solutions of the equations of
motion \cite{brack85,centelles90}.

It is known since long that the variational solution of $E_{HF}^{ETF}$ at $\hbar^2$ order
including the spin-orbit contribution overbinds the nuclei and gives neutron and proton 
radii too small and a diffuseness too sharp as compared with the corresponding quantal HF 
results \cite{brack85,centelles90}. This fact is probably due to the truncation of the 
semiclassical expansion at second order. To try to simulate higher-order contributions, at least
in the case of Skyrme interactions, Krivine and Treiner proposed to renormalize the $\hbar^2$ 
contribution to the kinetic energy \cite{treiner86}. Following this idea, we renormalize the 
$\hbar^2$ kinetic energy by a factor  $\beta_{KT}$ to obtain a semiclassical energy of 
the nucleus $^{208}$Pb $\sim$$-$1625 MeV, such that, by adding a shell correction $\sim$$-$11 MeV, one 
gets a total energy close to the experimental value  $\sim$$-$1636 MeV. This relatively small 
value of shell correction for $^{208}$Pb has been argued to be a robust ``experimental'' value 
(see, for example, \cite{MS.66,ZEL.67,BRA.72}) of the shell correction.

A very efficient way of recovering quantal effects, which are absent in semiclassical approximations 
of ETF type, is, in the spirit of the Kohn-Sham scheme, to replace in the semiclassical energy density 
the ETF particle, kinetic energy and spin densities by the corresponding HF 
expressions (see for instance  Refs.\cite{gridnev98,soubbotin00}). Applying now the variational 
principle using the single-particle wave-functions $\phi$ and $\phi^*$ for each type of particles
as functional variables, one can write the following set of coupled single-particle differential 
equations:
\begin{equation}
H_q \phi_i = \epsilon_i \phi_i,
\label{eq25}
\end{equation} 
where the single-particle Hamiltonian $H_q$ reads 
\begin{equation}
H_q = -{\bf \nabla}\frac{\hbar^2}{2m_q^*({\bf R})}{\bf \nabla} + U_q({\bf R})
-i{\bf W}_q({\bf R})\cdot \left[{\bf \nabla} \times {\bf \sigma}\right] \, ,
\label{eq26}
\end{equation}
where the effective mass, the mean-field and the spin-orbit potential for each kind of nucleons 
are given by
\begin{equation}
\frac{\hbar^2}{2m_q^*(r)} = \frac{\delta {\cal H}}{\delta \tau_q} \quad
U_q = \frac{\delta {\cal H}}{\delta \rho_q} \quad
{\bf W}_q({\bf r}) =  \frac{\delta {\cal H}}{\delta {\bf J}_q} \,.
\label{eq27}
\end{equation}

This HF method reported here was introduced formally in Ref.\cite{soubbotin03} through
a quasi-local reduction of the non-local density functional theory and generalized
later on to take into account pairing correlations \cite{krewald06}. 
Using this approximation one can solve the HF problem with finite-range interactions in 
coordinate space, if spherical symmetry is assumed, with the same simplicity as for the case 
of Skyrme forces. It is found that this reproduces extremely well the full HF or HFB calculations 
as it can be seen in Refs.~\cite{soubbotin03,krewald06,behera16}.

\section{Numerical investigations}
Our aim in this paper is twofold. On the one hand, we want to provide parametrized ETF 
densities, single-particle and spin-orbit potentials as well as the effective mass for both, 
neutron and protons, which average the corresponding HF values obtained by solving Eqs.~(\ref{eq25})-(\ref{eq27})
with the same interaction. To this end we use the D1S Gogny force \cite{berger91},
which has been extensively applied in different nuclear calculations (see, for example, \cite{d1s}). 
These fully quantal calculations at HF or HFB levels, however, are usually numerically expensive, 
because of the fact that the Gogny interactions are finite range interactions. Therefore, it is 
interesting and important to investigate the possibility of the existence of an accurate 
parametrization in terms of the mass and atomic numbers of different ETF quantities associated 
with the Gogny D1S interaction, which can be used  without performing explicitly the more cumbersome 
HF or HFB calculations. 

On the other hand, these parametrized single-particle potentials fitted in this 
work could be used profitably to compute the shell corrections and therefore to estimate,
perturbatively, the HF energy without performing explicitly the full HF calculation. In particular, 
the parametrized potentials are a key ingredient to obtain the shell-corrections through the 
so-called Wigner-Kirkwood $\hbar$-expansion method \cite{bhagwat10,bhagwat12} and this will be used 
in the Microscopic-Macroscopic calculation of nuclear masses based on the Gogny interaction reported 
in the paper II. It is worth to point out that this perturbative calculation of the 
shell corrections using semiclassical mean-field potentials has been recently applied to describe
the inner crust of neutron stars in the Wigner-Seitz approximation using Gogny forces \cite{mondal20}.  

\subsection{Neutron and proton densities}

\begin{figure*}[htb!]
\centering \subfigure [][~Neutron Density]{\centering \includegraphics[scale=0.37]{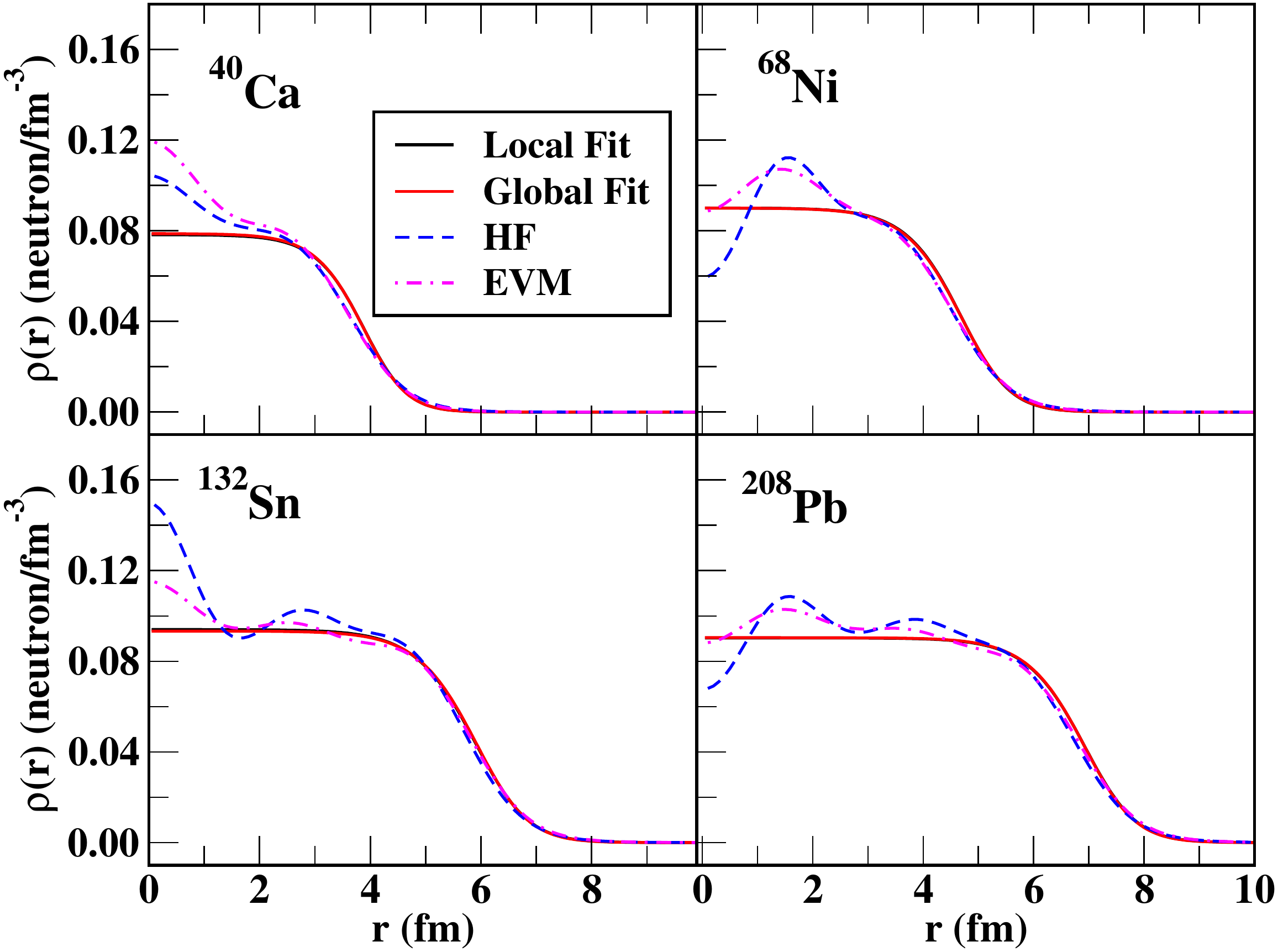}}
\centering \subfigure [][~Proton Density]{\centering \includegraphics[scale=0.37]{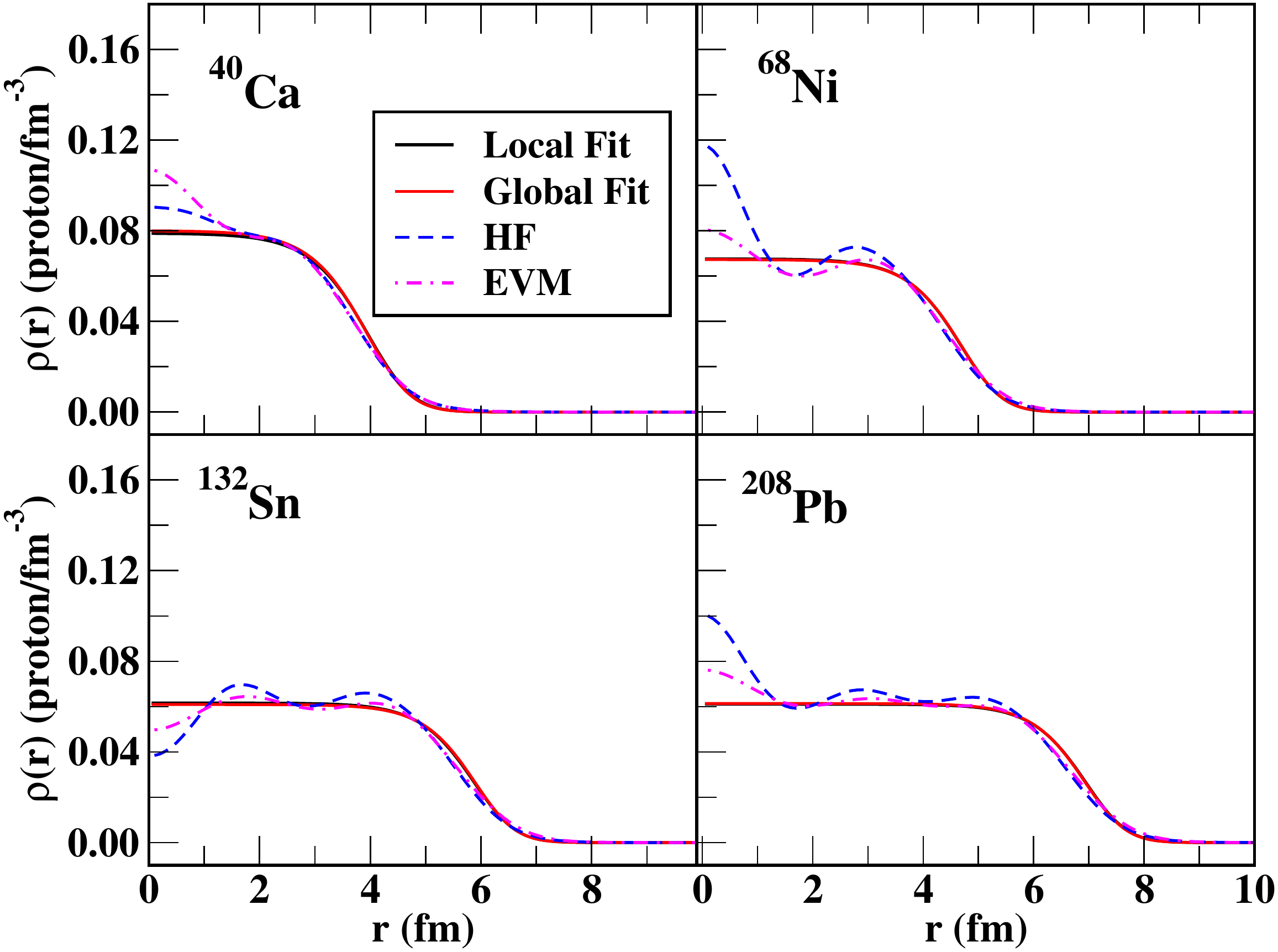}}
\caption{The nucleon density distributions for $^{40}$Ca, $^{68}$Ni, $^{132}$Sn 
and $^{208}$Pb. The densities obtained by using the local and global parameters
along with those from the Hartree-Fock calculations (denoted HF) and 
Expectation Value Method (denoted EVM, see section IV for further details) are 
displayed in the figure.}
\label{dens}
\end{figure*}

In this work the trial density profiles are chosen of the modified Fermi type. Explicitly,
\begin{eqnarray}
\rho\left(r\right) ~=~\rho_o\,g\left(r;R,a\right)
\end{eqnarray}
with
\begin{eqnarray}
g\left(r;R,a\right)~=~ f^{\nu}\left(r;R,a\right)
\label{eq20a}
\end{eqnarray}
and
\begin{eqnarray}
f\left(r;R,a\right) ~=~\frac{1}{1 + e^{\left(r - R\right)/a}},
\end{eqnarray}
where the half-density radius ($R$) and diffusivity ($a$) of each type of particles
are the variational parameters, and $\rho_o$ is to be determined by the norm requirement.
It is known that the exponent $\nu$ in (\ref{eq20a}) is practically unaffected 
by the minimization procedure \cite{brack85}, remaining always close to its starting value. 
Therefore, we fix values of these indices to be $\nu$ = 1.5 for neutrons and $\nu$ = 2.5 
for protons. The reason for this choice will be discussed below in Section III.B.1.

Using these trial densities we minimize the ETF-HF total binding energies
with a $\beta_{KT}$ factor equal to 1.9 and adding the direct and exchange
(at Slater level) Coulomb contributions for a large set of 551 even-even nuclei
(assumed to be spherical) covering a wide range of charge numbers, namely, from $^{20}$Ne to
$^{232}$Ds. 
With the aforementioned values of the exponents $\nu$ of the neutron and proton densities 
and of $\beta_{KT}$, the binding energy for $^{208}$Pb works out to be $\sim$1625 MeV, as discussed before.

As already mentioned, one of the aims of this work is to attempt a parametrization
of the ETF self-consistent densities, mean fields and spin-orbit form
factors computed with the Gogny D1S force, which give the average behaviour
of the corresponding HF densities and fields. This in turn implies a systematization
of the half-density radii and diffusivities as a function of $N$ and $Z$. The deformation
can then be incorporated within this scheme by appropriately defining the distance
function (see, for example, \cite{bhagwat10,bhagwat12}). Thus, in order
to study the systematics of the Fermi distribution parameters,
spherical symmetry is assumed in the ETF calculations. The half-density
radii and diffusivities thus obtained for the individual nucleus
through a minimization process will be generically called `local' parameters henceforth.
The systematized radii and diffusivities as functions of $N$ and $Z$ will be 
termed as `global' parameters. The corresponding results in the figures below have 
been labeled by the terms `local fit' and `global fit', respectively. 

The kinetic energy densities, mean fields, spin-orbit potentials and the effective 
masses are obtained once the density parameters are optimized through a minimization 
process. These quantities will be termed as `exact'. These are then fitted to 
model form factors of the modified Fermi type for the each individual nucleus. The parameters thus obtained for the
individual nucleus will be termed `local' parameters, whereas the systematized radii 
and diffusivities as functions of $N$ and $Z$ will be termed as `global' 
parameters. As in the case of the densities, the quantities obtained by using these 
parameters will be called `local fit' and `global fit' respectively. 

The optimized half-density radii and diffusivity parameters for each kind of 
densities are modeled as
\begin{eqnarray}
R^{(i)} &=& r_{1}^{(i)}\left(1 + r_{2}^{(i)}\,I\right)A^{1/3} + r_{3}^{(i)} \\
a^{(i)} &=& a_{1}^{(i)}\left(1 + a_{2}^{(i)}\,I\right)
\end{eqnarray}
where, $i$ = n (neutrons) or p (protons), $I =\left(N-Z\right)/A$ is the asymmetry 
parameter, and $r_{1,2,3}^{(i)}$ and $a_{1,2}^{(i)}$ are free parameters of the model. 
These are determined by using an implementation of the well-known Levenberg - Marquardt 
algorithm \cite{MAR.68,KEL.99}. The fits work out to be excellent, with {\it rms} 
residues being of the order of or even smaller than 10$^{-2}$ in all the cases. The 
explicit values of these parameters have been listed in Appendix 2.

In Figures 1(a) and 1(b) we display the point neutron and proton densities 
obtained from the minimization procedure discussed previously for the nuclei 
$^{40}$Ca, $^{68}$Ni, $^{132}$Sn and $^{208}$Pb (`local fit') together with the densities 
provided by the global fit in comparison with the HF densities obtained by solving the
set of Schr\"odinger equations (Eq. (\ref{eq25})).
From these figures we also can see that the semiclassical densities 
nicely average the quantal oscillations and have a similar fall-off at the surface.

\begin{figure*}[htb]
\centering \includegraphics[scale=0.31]{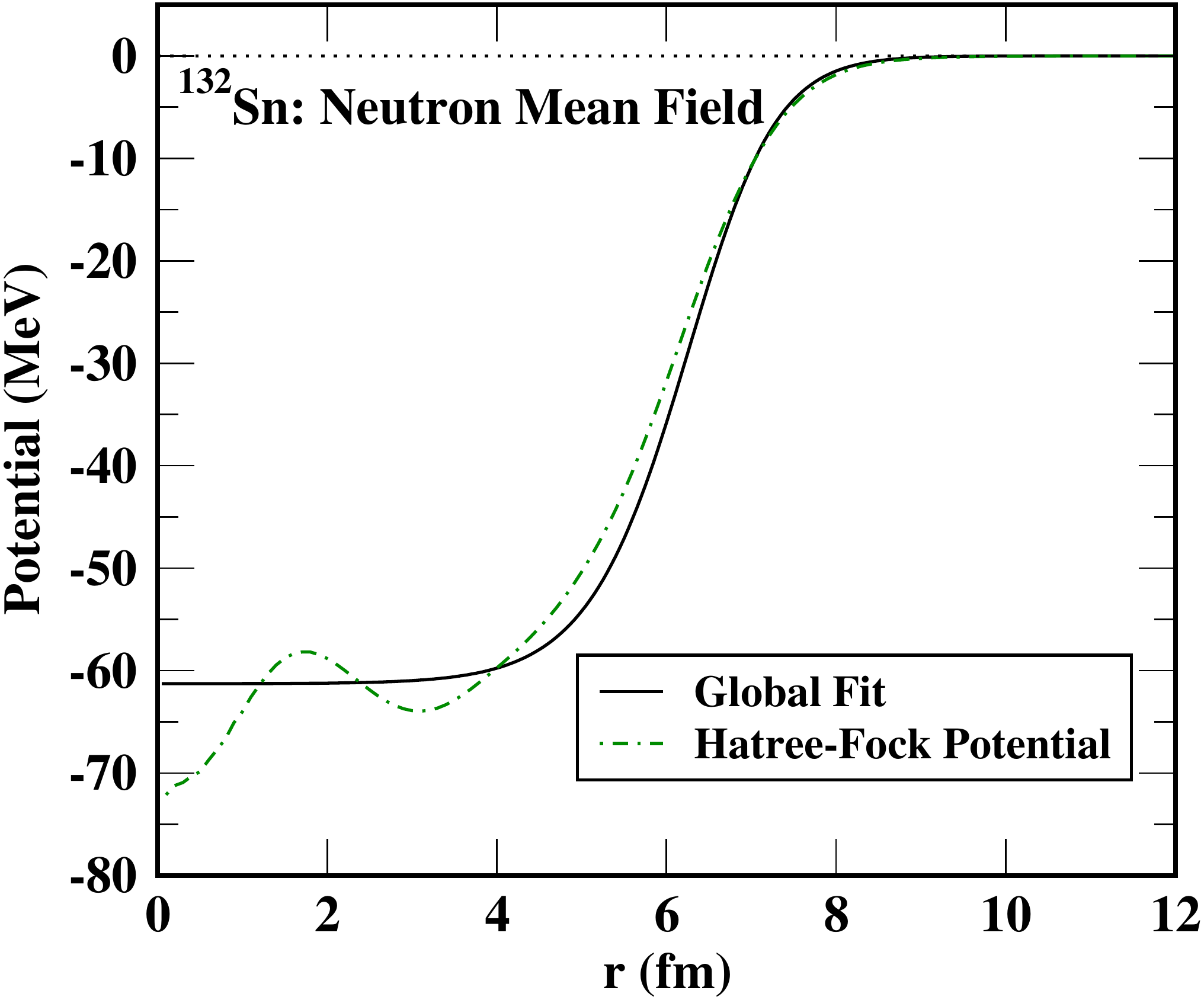} 
           \includegraphics[scale=0.31]{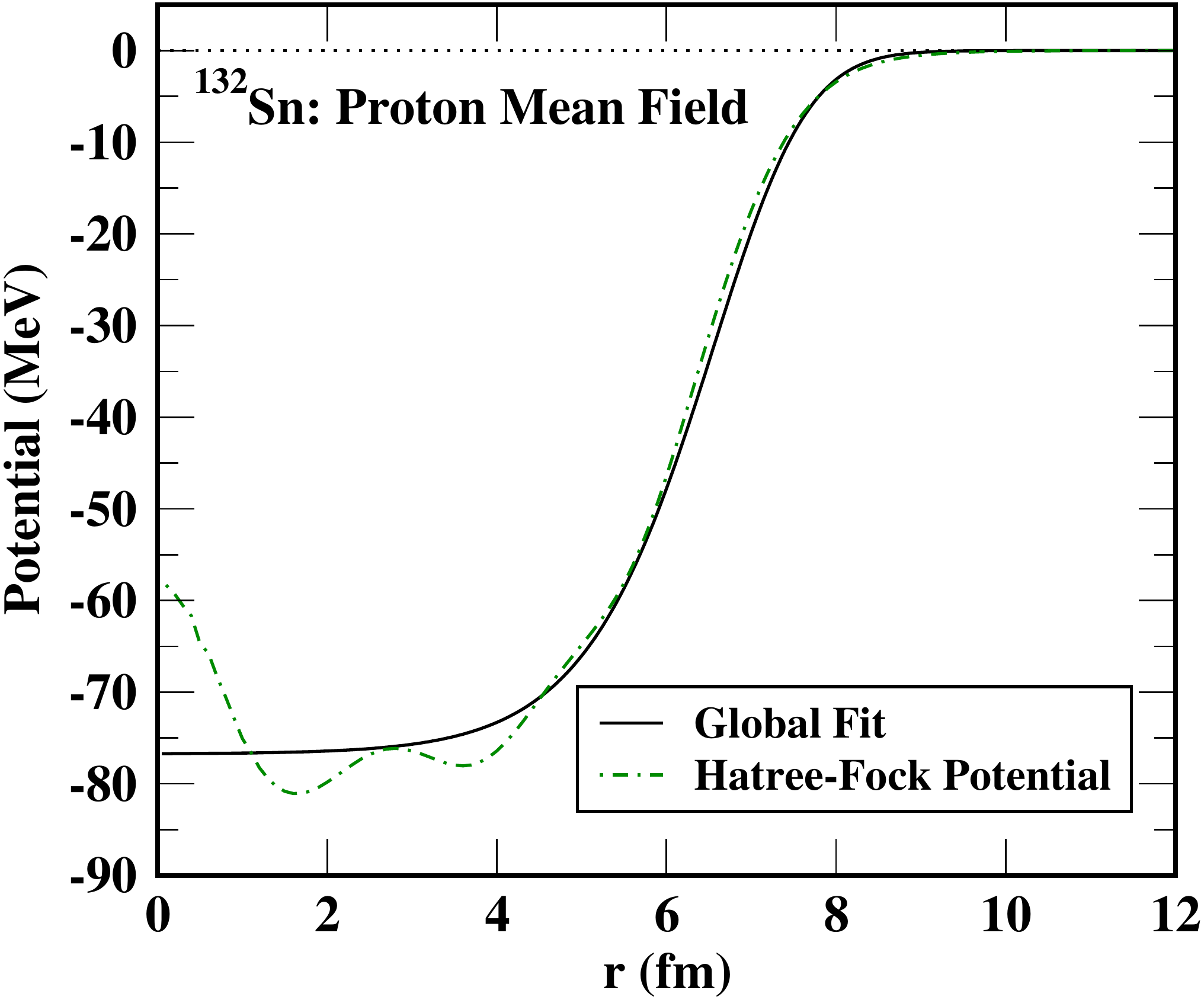}
\\
\centering \includegraphics[scale=0.31]{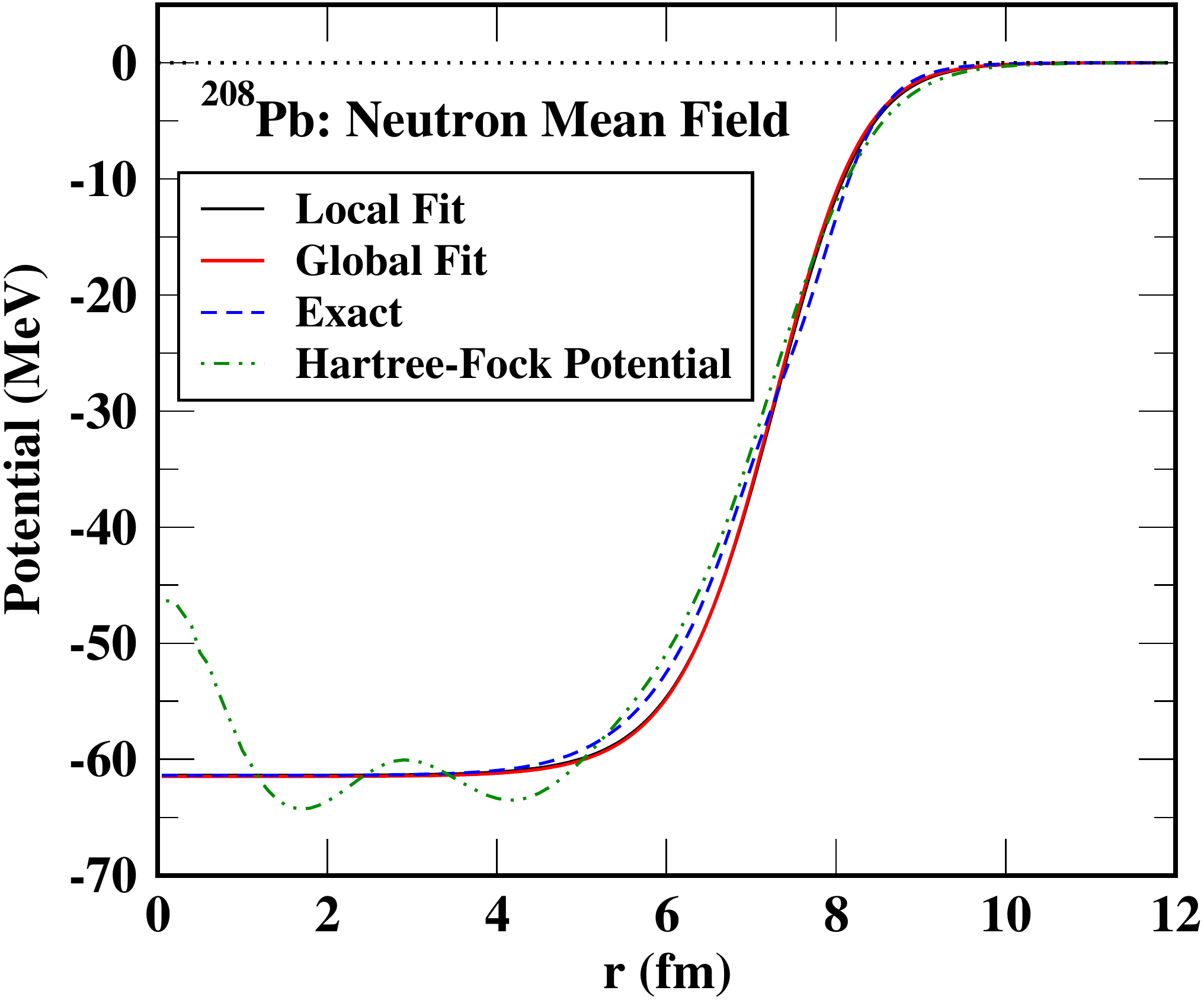} 
           \includegraphics[scale=0.31]{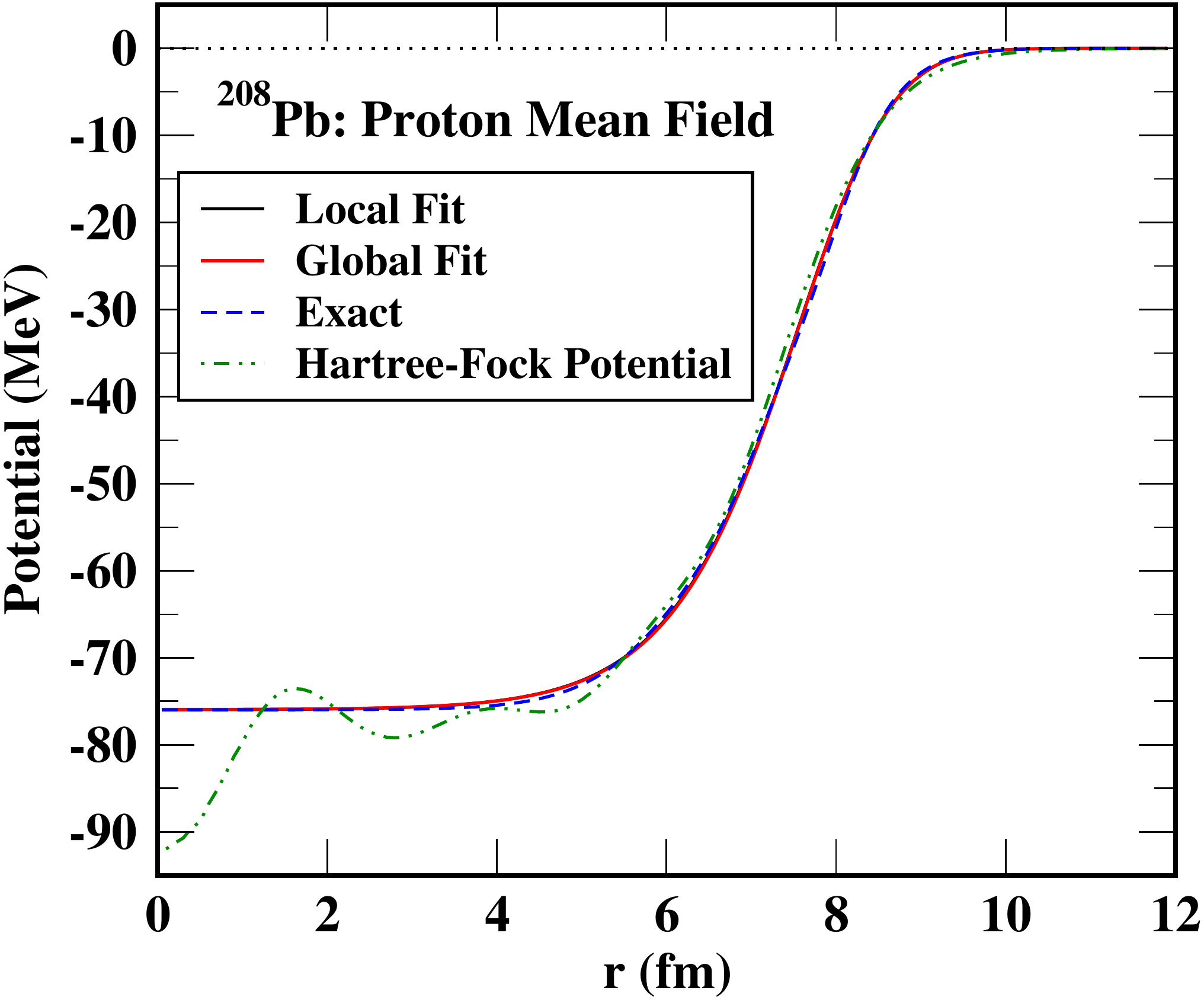}
\caption{Calculated and fitted mean fields for $^{132}$Sn and $^{208}$Pb compared with
the corresponding HF potentials.}
\label{pot}
\end{figure*}

\begin{figure*}[htb]
\centering \includegraphics[scale=0.31]{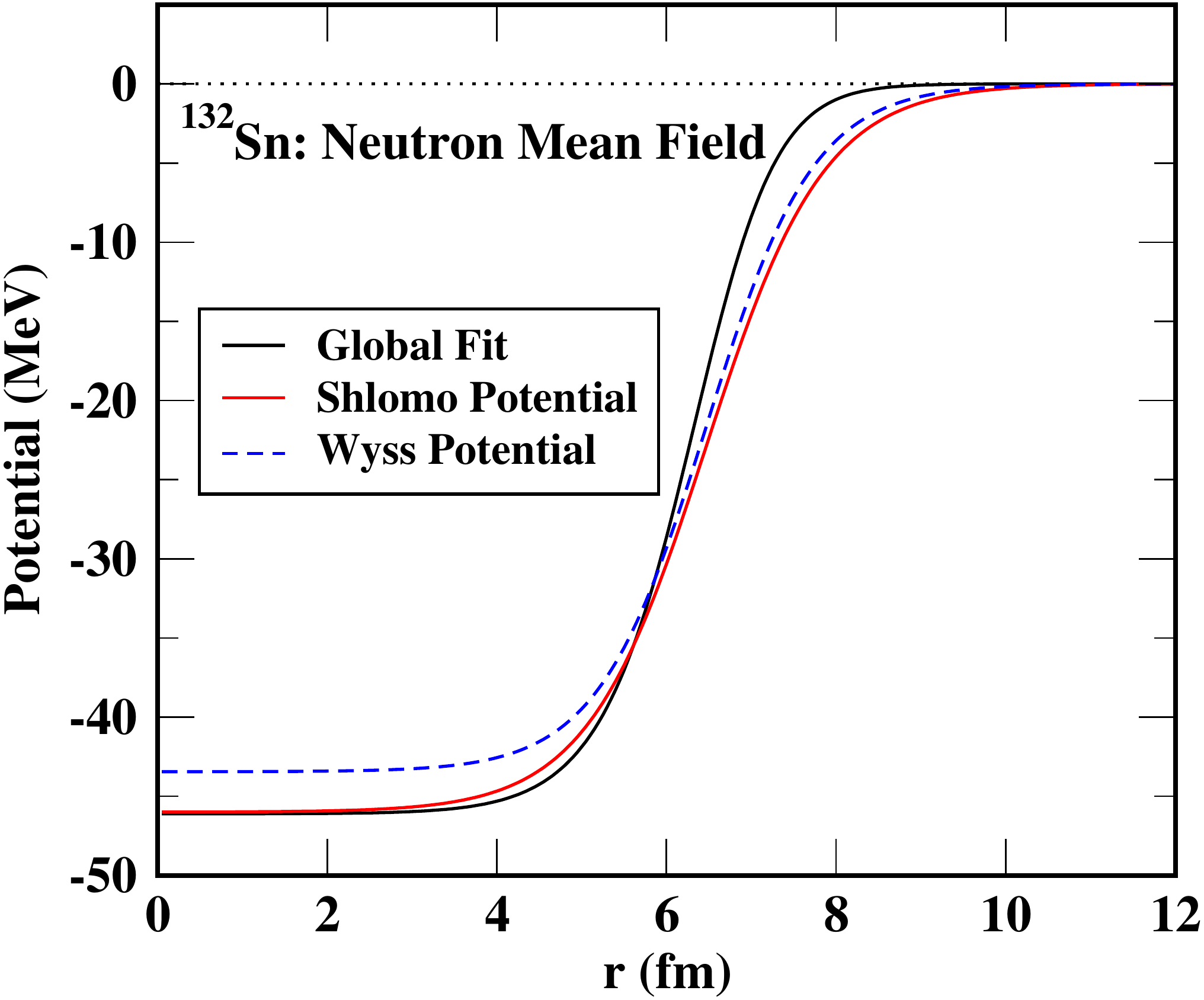} 
           \includegraphics[scale=0.31]{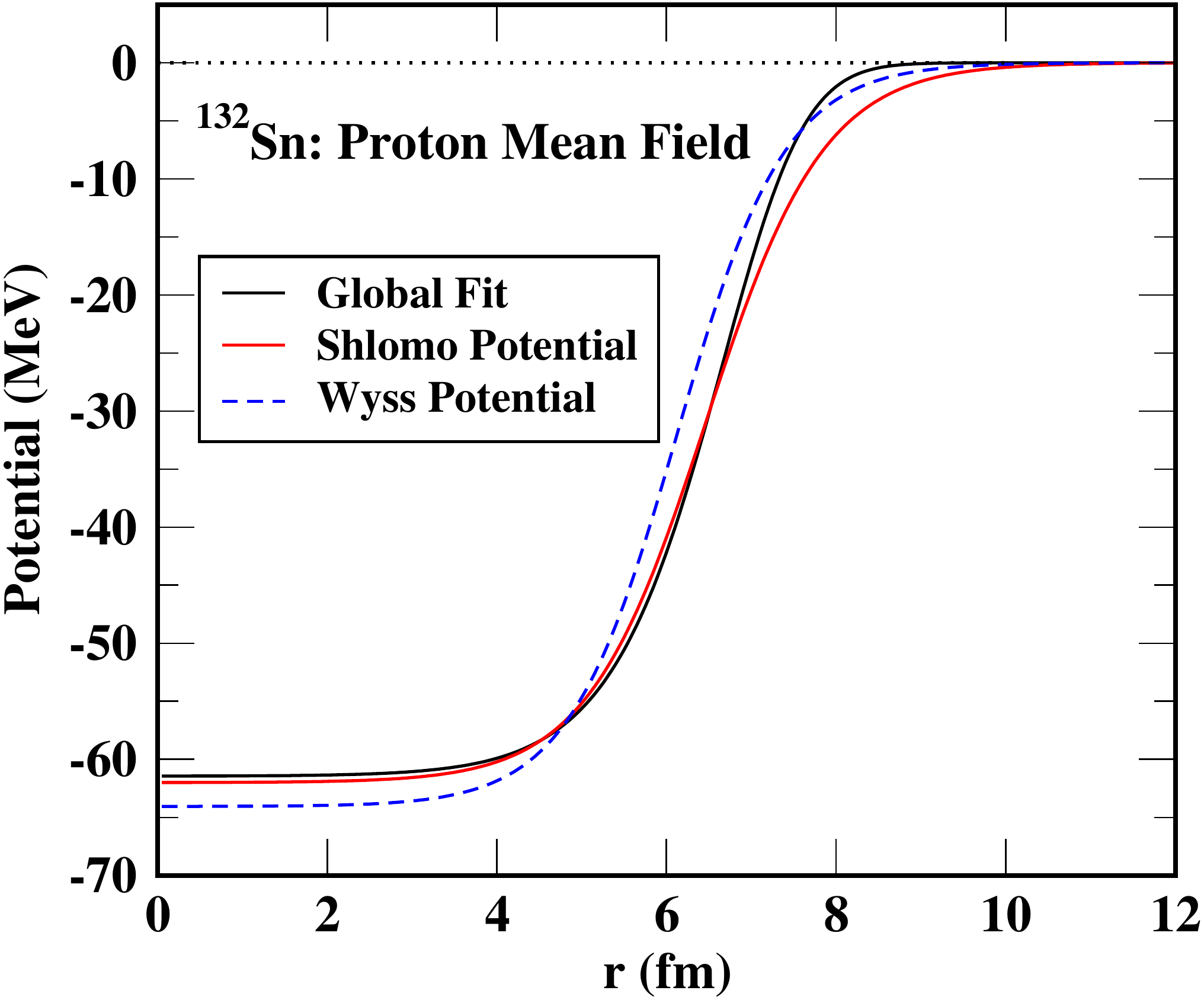}
\\
\centering \includegraphics[scale=0.31]{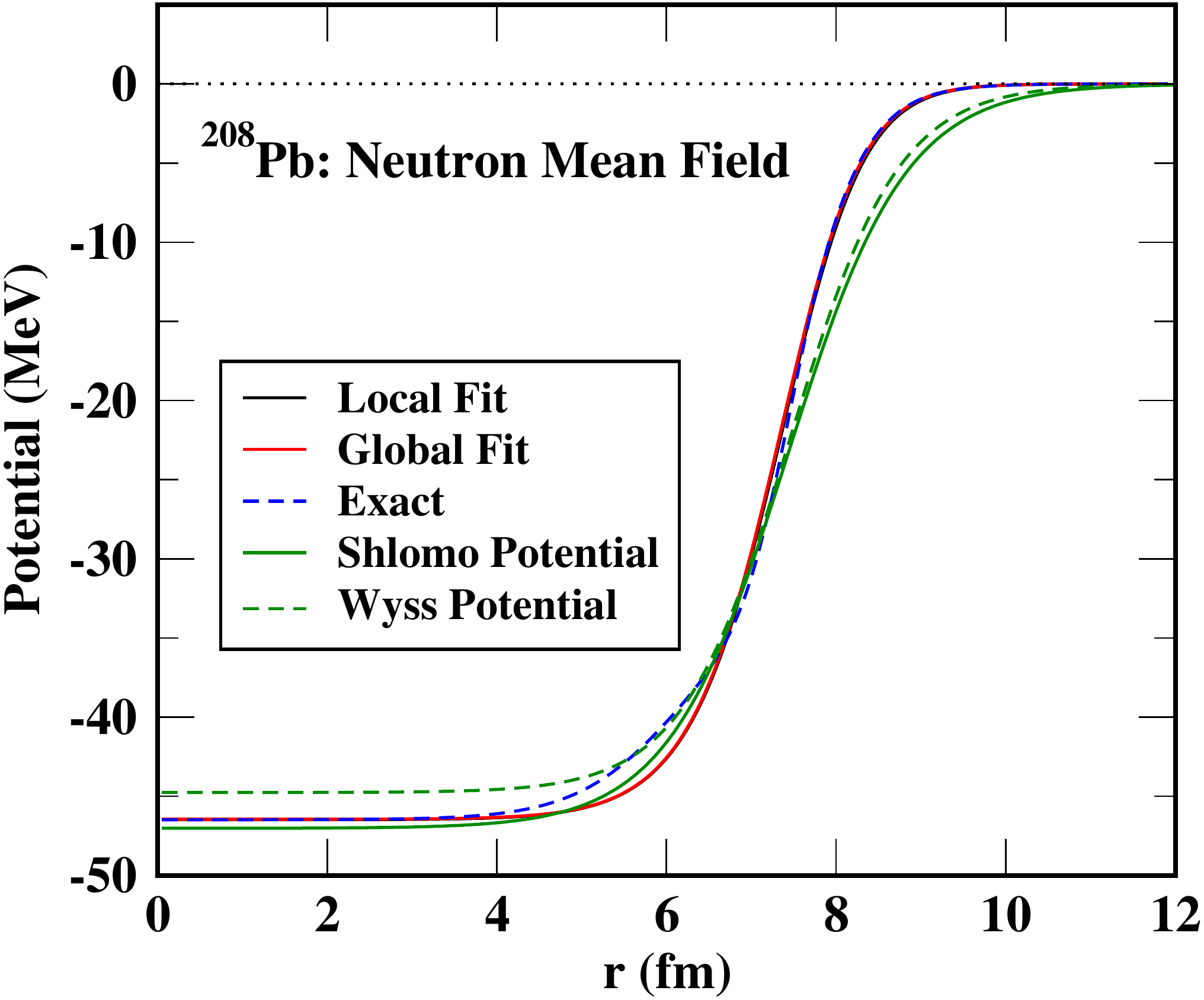} 
           \includegraphics[scale=0.31]{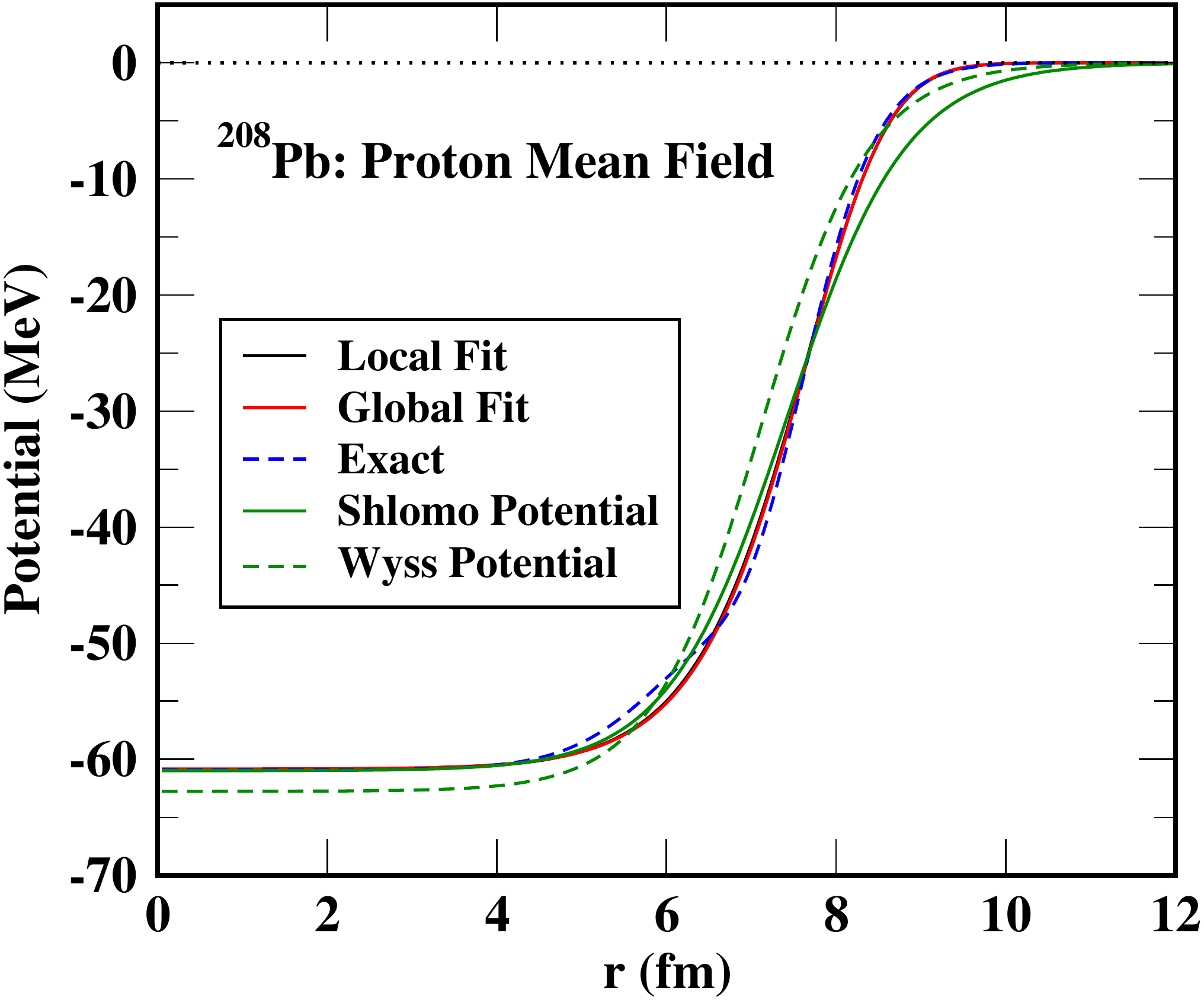}
\caption{Calculated and fitted single-particle potentials with $m^*=m$ for 
$^{132}$Sn and $^{208}$Pb compared with the phenomenological mean-fields of 
	Refs.\cite{bhagwat10,bhagwat12} and \cite{shlomo92}.}
\label{pheno-pot}
\end{figure*}

\subsection{Parametrized HF potentials and effective masses}

We next investigate the possibility to parametrize the smoothly varying part of the HF neutron 
and proton mean-field, the spin-orbit potentials and the $r$-dependent effective masses 
employing the semiclassical ETF model described above. At HF level, the single-particle 
potential for each type of particles is obtained as the functional derivative 
of the energy density (Eq. (\ref{eq2})) with respect to the corresponding particle density (Eq. (\ref{eq27})). 
The corresponding ETF mean-field potentials, which we call ETF-HF potentials, 
are obtained by replacing in the HF potential the quantal kinetic energy and spin densities 
by the corresponding ETF counterparts, Eqs.(\ref{eq8}) and (\ref{eq9}), respectively. In an 
equivalent way, the semiclassical mean-field potential can also be derived from the ETF 
energy density functional using explicitly Eq.(\ref{eq24}) of the Appendix and performing 
the variation with respect to the corresponding type of density, keeping the kinetic energy 
and spin densities as independent variables, as in the quantal case,  and replacing them by 
the ETF expressions after variation.

Using the variational densities we can obtain all the quantities derived from the energy density 
functional, as for example the mean-field or the effective mass for each type of nucleons (which 
are the so-called `exact' quantities discussed in the previous subsection).
However, these `exact' quantities are obtained numerically and are not easy to handle in a global way.
We therefore undertake a second fit procedure, where we directly express these  quantities by 
modified Woods-Saxon form factors whose parameters are adjusted to reproduce the `exact' results. 
The parameters obtained from this local fit for a large set of nuclei are finally expressed as a 
function of the mass and atomic numbers, which allow to express the nuclear mean-fields, spin-orbit 
potentials and effective masses in a global form, which is ready for an easy and direct use as 
were before the phenomenological Woods-Saxon potentials (without effective mass) as given, e.g., 
by Shlomo \cite{shlomo92}.
 
\subsubsection{Mean Field}
Let us start with the mean-field. Specifically, we use:
\begin{eqnarray} 
V\left(r\right) &=& V_{o} f^{\nu}\left(r;R,a\right)
\end{eqnarray} 
where the symbols have the same meaning as before. The indices $\nu$ for the neutronic
and protonic mean fields have been taken to be 1.5 and 3.0 
respectively.
In the global fit the strengths, diffusivities and half-density radii are parametrized as: 
\begin{eqnarray} 
V_{m}^{(i)} &=& V_{1}^{(i)}\left(1 + V_{2}^{(i)}\,I\right) + V_{3}^{(i)}A^{1/3}\\
a_{m}^{(i)} &=& b_{1}^{(i)}\left(1 + b_{2}^{(i)}\,I\right) \\
R_{m}^{(i)} &=& z_{1}^{(i)}\left(1 + z_{2}^{(i)}\,I\right)A^{1/3} + z_{3}^{(i)}
\end{eqnarray} 
As before, the index $i$ stands for neutrons (n) and protons (p). The quantities 
$V_{k}$, $b_{k}$ and $z_{k}$ are free parameters. 

Here, we carry out two different fits of the neutron and proton single-particle potentials.
In the first case, the optimal densities, deduced through the variational 
procedure discussed before, are used to obtain the single-particle HF potentials (ETF-HF 
potentials) for neutrons and protons as explained before. Those potentials are then
fitted to modified Woods-Saxon form factors. These potentials can be directly compared 
with the true HF potentials obtained self-consistently. 

In the second case, the contribution from 
the $r$-dependent effective mass is subsumed in the potential energy, thereby 
ensuring that only the constant ($r$-independent) bare mass appears in the kinetic energy 
operator. The motivation for this approach comes from the fact that the optical
potentials, the phenomenological potentials used in the Mic-Mac calculations of nuclear masses, as 
well as the potentials fitted to reproduce the experimental observed single-particle
states available in the literature have the effective mass equal to the bare mass. 
It should also be mentioned that the existing programs for this type of potentials work with 
bare mass and then would be complicated to include the effective  mass in them, in particular
in the deformed case. A few examples being, the potential proposed by Shlomo \cite{shlomo92}
and the Wyss potential, used in our previous Mic-Mac calculations 
reported in Refs.\cite{bhagwat10,bhagwat12}.
This scenario can also be easily simulated with our ETF approach. 
Notice that in the ETF energy density the contribution (Eq. (\ref{eq24})) can be 
split into a pure kinetic energy term with the bare mass (Eq. (\ref{eq11})) and a 
$\hbar^2$ contribution to the exchange energy (Eq. (\ref{eq17})). In this way, 
the effective mass contributions to the energy density have been subsumed in the 
potential part. This ensures that only the bare mass appears in the kinetic 
energy term (Eq. (\ref{eq11})) and that the total ETF energy remains unchanged.
Within the strict semiclassical ETF approximation, the kinetic energy densities entering in 
the $\hbar^2$ exchange energy (Eq. (\ref{eq17})) have to be replaced by their semiclassical counterparts
(Eq. (\ref{eq8})), which in turn are functionals of the particle densities. Only then one computes
the neutron and proton single-particle potentials as functional derivatives of the potential energy
density with respect to the density of each type of particles. We refer to the neutron and proton 
single-particle potentials  obtained in this way as ETF phenomenological mean-field  potentials. 

For each single-particle potential, the strengths, the half-density radii and 
the diffusivities are systematized using the ansatz defined earlier (Eqs. (24)-(26)).
The optimization of the parameters $V_{k}$, $b_{k}$ and $z_{k}$ is done separately in the 
two cases, by employing again the Levenberg-Marquardt algorithm. In both these cases,
the fits are found to be good, with {\it rms} residues of the order of 10$^{-2}$. The 
explicit values of the parameters appearing above in both cases are listed in Appendix 2
under ``Hartree-Fock mean field potentials'' and ``Phenomenological mean field potentials'',
respectively. It is worth to remember here that the ``Exact" mean-field potentials are 
complicated functions of the proton and neutron densities, which in turn depend on the exponent $\nu$. 
The ``Exact" potentials are then fitted to generalized Fermi functions. We have chosen the exponents $\nu$
in the neutron and proton densities and mean-fields by a double optimization procedure in such a way that 
the fitted, local or global, single-particle potential for each type of particles reproduces as well as 
possible the ``Exact" mean field. The fact that the proton mean field is slightly flatter at the 
bottom than the neutron mean field (see Figs.~\ref{pot} and \ref{pheno-pot}) is the reason why the proton
exponents, for both density and mean field, are larger than the corresponding neutron exponents.

\begin{figure*}[htb]
\centering \includegraphics[scale=0.31]{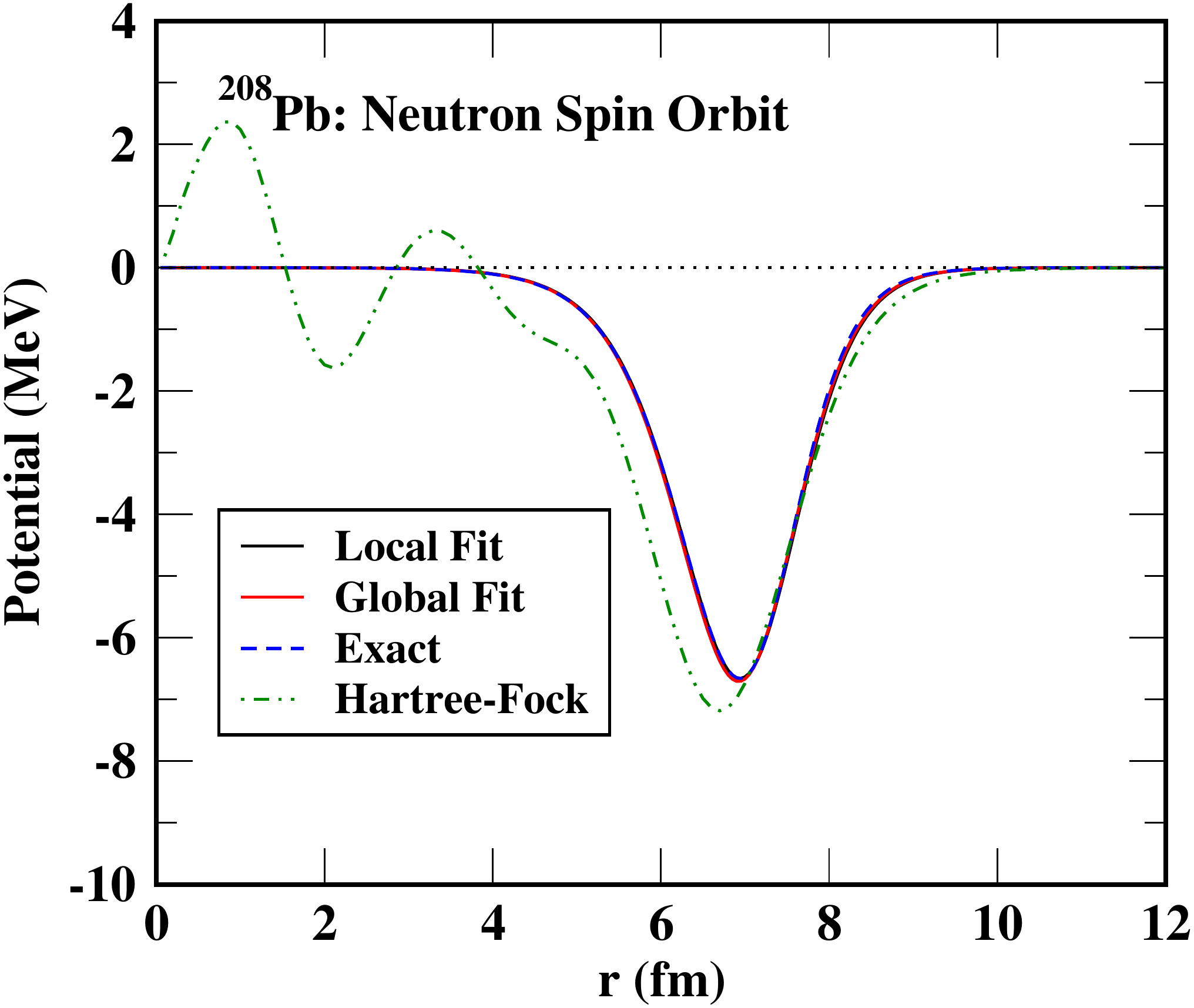}
           \includegraphics[scale=0.31]{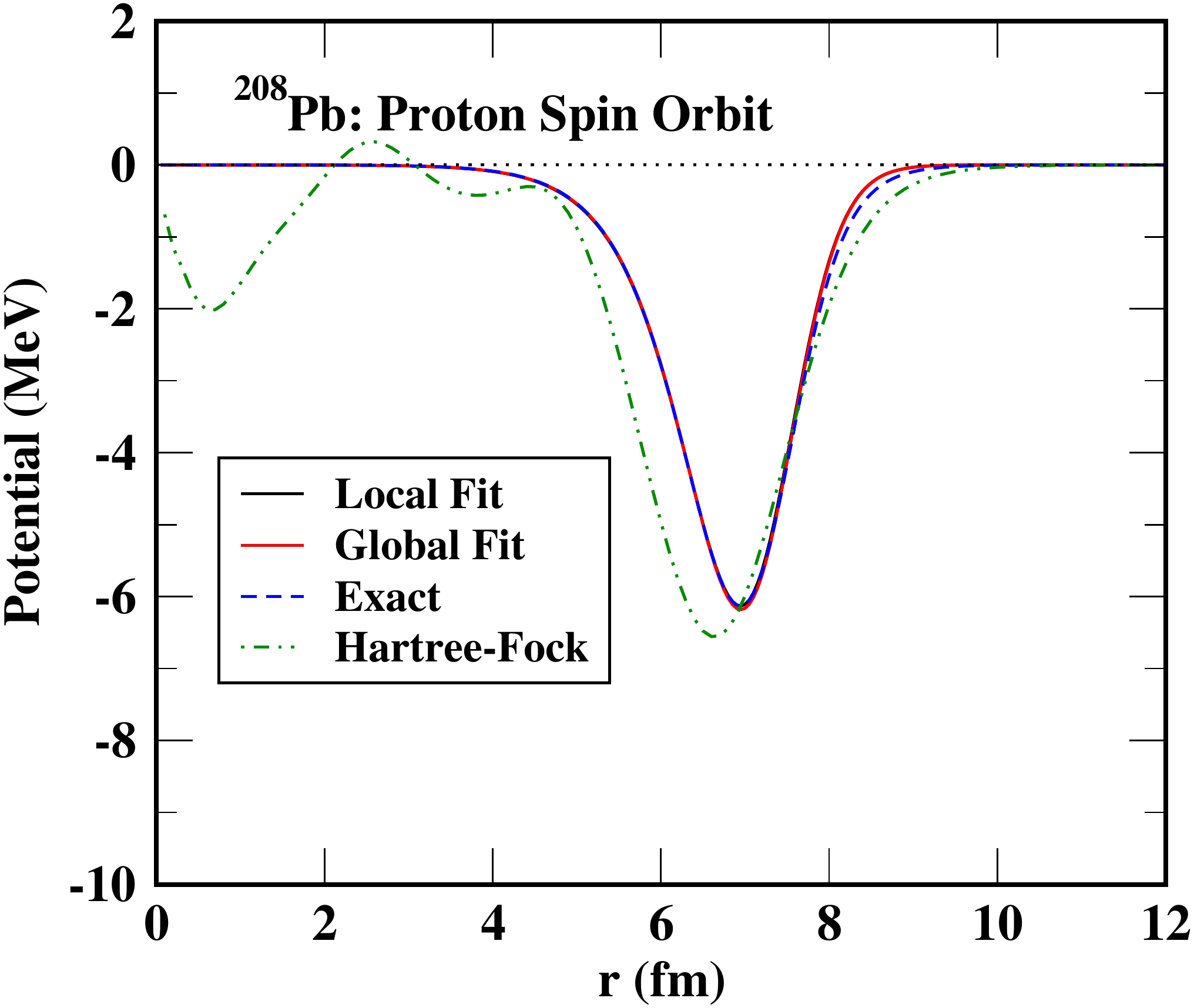}
\caption{Calculated and fitted SO potentials for $^{208}$Pb compared with the corresponding 
HF results.}
\label{so}
\end{figure*}

\begin{figure*}[htb]
\centering \includegraphics[scale=0.31]{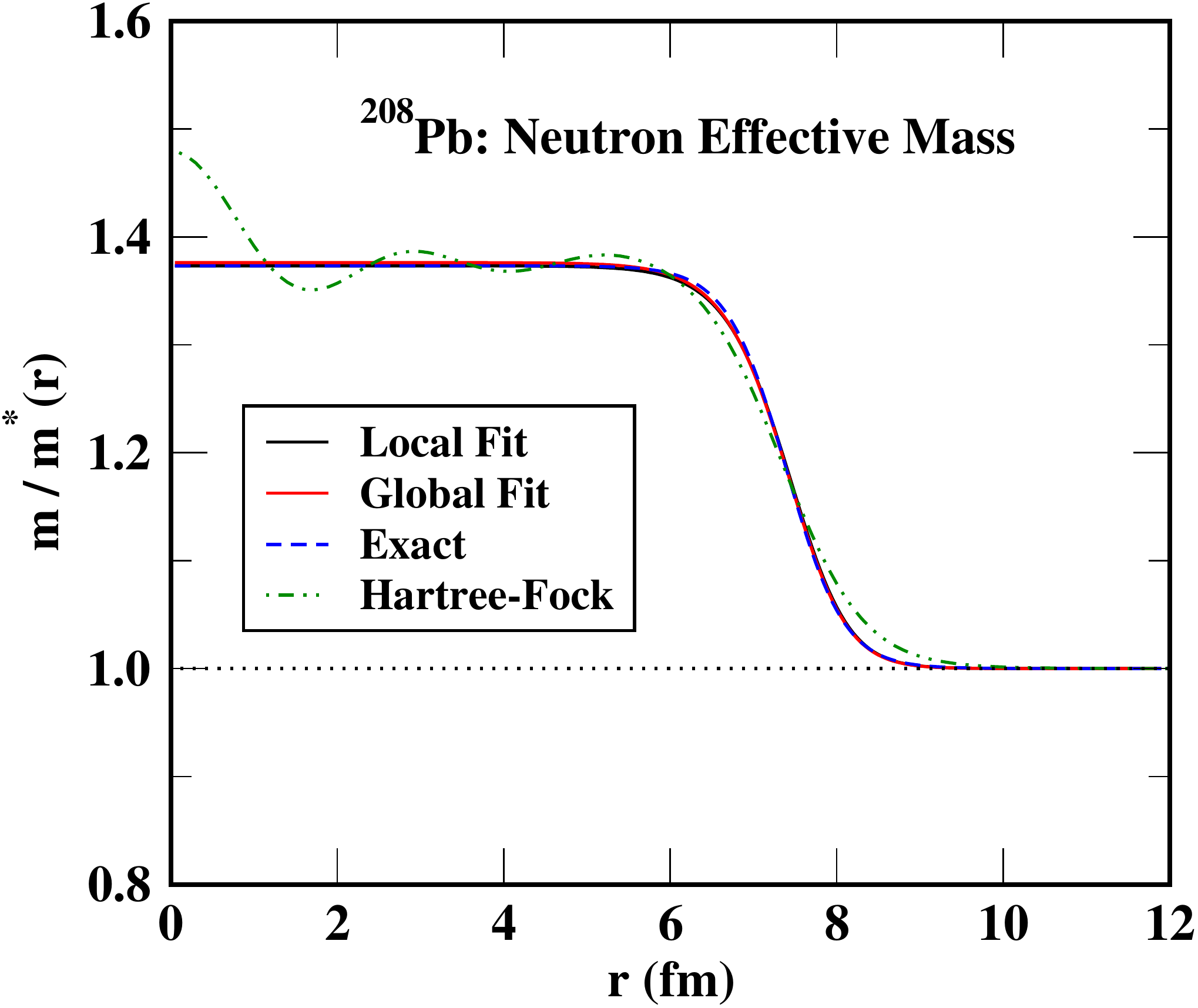}
           \includegraphics[scale=0.31]{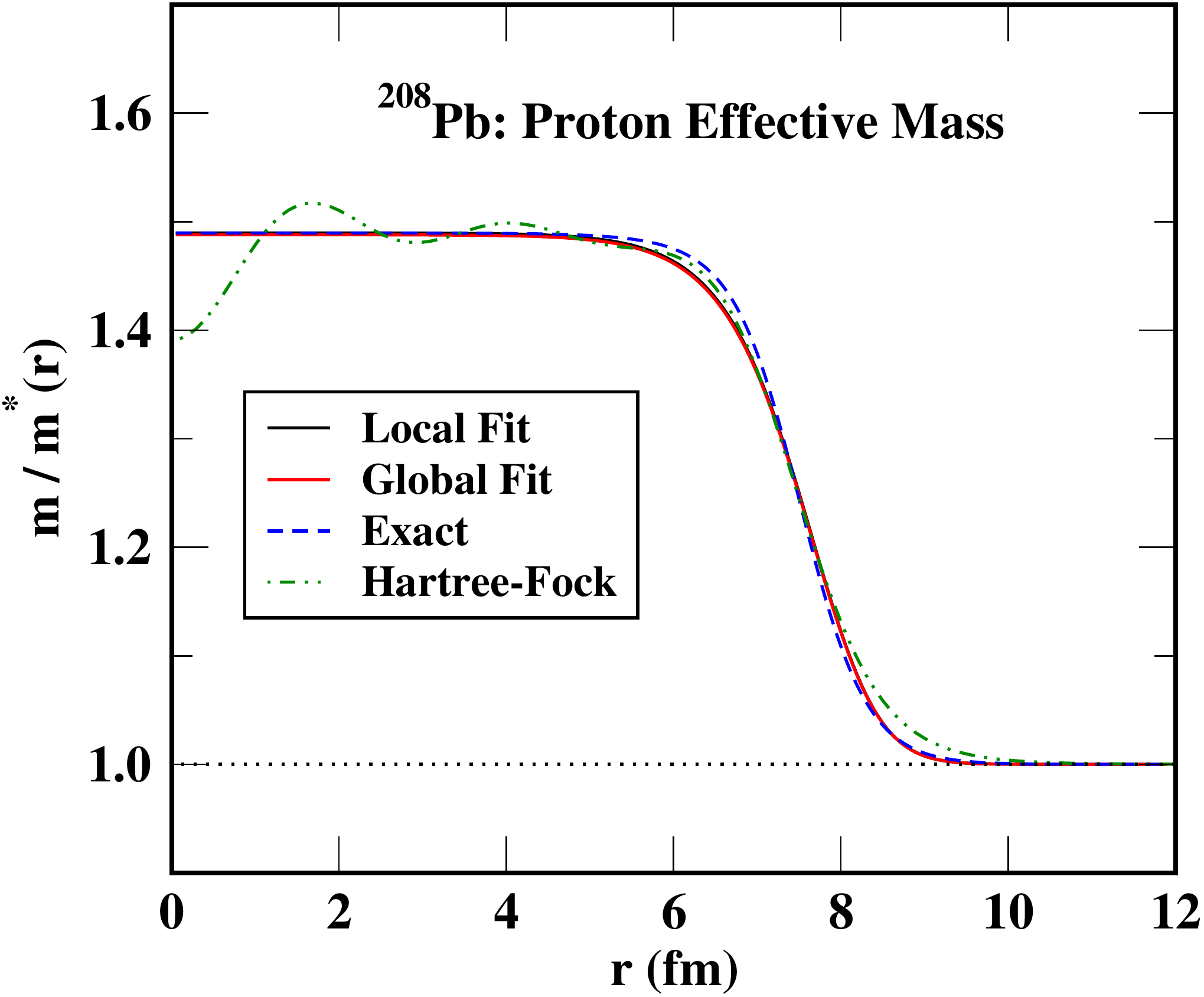}
\caption{Calculated and fitted effective masses for $^{208}$Pb compared with the corresponding 
HF results.} 
\label{emm}
\end{figure*}

As an example, the actual ETF-HF mean fields obtained using the variational densities (denoted by `exact'), 
as well as those obtained from the local and global parametrizations are plotted in the lower panel 
of Figure \ref{pot} for $^{208}$Pb. In the upper panel of the same figure we also show the `global fit' 
and HF potentials for the nucleus $^{132}$Sn. The single-particle potentials with $m^{*}=m$ are 
displayed in Figure (\ref{pheno-pot}). In the latter case, the phenomenological mean fields 
used in Refs. \cite{bhagwat10,bhagwat12}, labeled `Wyss Potential' as well as the one 
reported by Shlomo \cite{shlomo92}, labeled `Shlomo Potential', are also plotted for 
comparison. In the former case, as can be seen in Figure \ref{pot}, the computed potentials 
nicely average out the HF potentials, which is as expected. Further, an excellent agreement between the 
different potentials for protons as well as neutrons is very clear from the above figure, indicating 
reliability of the fits achieved in this work. Regarding the phenomenological mean-field potentials, we see in 
the lower panel of Figure~\ref{pheno-pot}, that both, the local and the global fits, nicely reproduce 
the exact potential. In the upper panel of this Figure we show that for the nucleus $^{132}$Sn 
the global fit of the ETF Gogny potential in the $m=m^*$ case qualitatively follows the trends of the 
Shlomo \cite{shlomo92} and Wyss \cite{bhagwat10,bhagwat12} phenomenological potentials. Similar results are 
found and the same comments hold for all the nuclei considered here. Actually, since the phenomenological mean 
field potentials of Refs.~\cite{shlomo92} and \cite{bhagwat10,bhagwat12} are fitted to experiment, 
one can assume that a finite effective mass is also implicitly 
included in those potentials. Therefore, the closeness of the phenomenological potentials with the one from the 
Gogny force where we incorporate the effective mass into the local mean-field potential is not so astonishing.  

\subsubsection{Spin-Orbit Potential}
The spin-orbit (SO) potential is studied next. 
Its form factor is given by Eq. (\ref{eq10}), which we parametrize as
\begin{eqnarray} 
V_{so}\left(r\right) &=& \frac{\nu U_{so}}{a}\,g\left(f - 1\right)
\end{eqnarray} 
where, $f$ and $g$ are as defined above, and the other symbols have their usual
meanings. Since the spin-orbit form factor (Eq. (\ref{eq10})) has been obtained by 
differentiating the neutron and proton densities, the indices $\nu$ for the spin 
orbit form factor have been chosen to be the same as those for the densities, namely, 
1.5 and 2.5 for neutrons and protons, respectively. 

The strengths, the diffusivities and the half-density radii of the
spin-orbit potentials are parametrized as:
\begin{eqnarray} 
U_{so}^{(i)} &=& U_{1}^{(i)}\left(1 + U_{2}^{(i)}\,I\right) + U_{3}^{(i)}A^{1/3}\\
a_{so}^{(i)} &=& b_{1}^{(i)}\left(1 + b_{2}^{(i)}\,I\right) \\
R_{so}^{(i)} &=& z_{1}^{(i)}\left(1 + z_{2}^{(i)}\,I\right)A^{1/3}
\end{eqnarray} 
As an example, the neutron and proton SO potentials obtained from Eq.(\ref{eq10}) 
using the variational densities ('Exact' calculation) for the nucleus $^{208}$Pb as 
well as the corresponding local and global fits are displayed in Figure~\ref{so}.
The excellent agreement between the different potentials for protons as well as
neutrons is very clear from the above figure. This observation for spin-orbit
potentials is particularly important, since the accurate description of spin-orbit
potential has a bearing on the ground-state masses as well.

For comparison, the HF values of the SO potentials are also plotted in the same
graphs. The two profiles are quite similar to each other, but it can be seen 
that the SO potentials obtained from HF calculations have smaller radii, and
they are a bit wider than the other potentials. This can be easily understood 
in the light of the density profiles, plotted in Figure (1). Notice that the
HF and ETF densities have very similar slopes, but the HF densities tend to 
bulge slightly more in the surface region. Spin-orbit potential being proportional
to the first derivative of the density profile, the ones obtained from HF and 
ETF densities are bound to be slightly different from each other, as is 
seen from Figure (1). It is important to point our that in order to 
reproduce the HF strength at the peak of the SO potentials, the spin-orbit strength
has been taken $W_0=-105$ MeV instead of the original D1S value $W_0=-130$ MeV. This 
is due to the fact that although the ETF densities are in good agreement with the HF
ones at the surface, their slopes are not exactly identical,  which has a direct impact
on the SO potential. As far as shell effects are dominated by the SO potential, we have
renormalized the ETF prediction to remain as close as possible to the HF result for the SO potential.

\subsubsection{Effective Mass}

We next assume that the inverse of the effective mass (Eq. (\ref{eq8})) computed at 
$k=k_{F_q}(R)$ can be parametrized as:

\begin{eqnarray} 
\frac{m}{m^{*}(r)} - 1 = M_{o}\,g\left(r;R,a\right)
\end{eqnarray} 
Here, the indices $\nu$ are chosen to be 1.5 and 2.5 respectively for 
neutrons and protons (same as those for densities). The strengths, half-density 
radii and diffusivity constants are parametrized as:
\begin{eqnarray} 
M_{o}^{(i)} &=& M_{1}^{(i)}\left(1 + M_{i}^{(n)}\,I\right) \\
a_{em}^{(i)} &=& c_{1}^{(i)}\left(1 + c_{2}^{(i)}\,I\right) \\
R_{em}^{(i)} &=& d_{1}^{(i)}\left(1 + d_{2}^{(i)}\,I\right) + d_{3}^{(i)}
\end{eqnarray} 
As in the case of the other quantities described above, the parameters
$M_{1,2}$, $c_{1,2}$ and $d_{1,2,3}$ are taken to be free parameters, and 
are determined through a $\chi^2$ minimization process. As observed for the parameters
of the mean field and the other quantities, the fits work out to be excellent,
with typical {\it rms} deviation being of the order of 10$^{-2}$ or smaller. 
The explicit values of these parameters can be found in Appendix 2. 

In Figure \ref{emm} we plot the quantity $m/m^{*}(r)$ for $^{208}$Pb, computed 
at ETF level using the variational densities
(denoted by `Exact'), the corresponding local and global fits as well as the HF result.
The results from the three approaches are almost indistinguishable 
from each other, indicating the good quality of the fits.
It can be seen that the HF and ETF $m/m^{*}(r)$ ratios are very similar to each
other. In fact, the ETF inverse effective mass nicely averages out that obtained from 
the HF calculations, as expected.

\section{Energy calculations}

\begin{table}[htb]
\begin{center}
\caption{Total ground-state energies obtained by using the Hartree-Fock, Expectation Value Method 
and ETF approaches.} \vspace{5pt}
\label{tab2}
\begin{tabular}{c|ccc} \hline
                   &  \multicolumn{3}{c}{Approach}    \\ \cline{2-4} 
 Nucleus           &    HF     &    EVM    &   ETF    \\ \hline
   $^{16}$O        &  -127.43  & -126.08   & -118.16  \\ 
   $^{40}$Ca       &  -341.70  & -339.34   & -338.95  \\ 
   $^{48}$Ca       &  -414.96  & -413.57   & -412.33  \\
   $^{56}$Ni       &  -479.82  & -476.36   & -479.01  \\
   $^{78}$Ni       &  -638.36  & -636.12   & -632.27  \\
   $^{90}$Zr       &  -783.51  & -781.63   & -785.75  \\
   $^{100}$Sn      &  -826.62  & -822.66   & -820.03  \\
   $^{132}$Sn      & -1100.72  &-1097.68   &-1089.09  \\
   $^{208}$Pb      & -1636.08  &-1632.37   &-1625.76  \\ \hline
\end{tabular}
\end{center}
\end{table}

\begin{table}[htb]
\begin{center}
\caption{The {\it rms} neutron and proton radii obtained by using the Hartree-Fock, Expectation Value Method 
and ETF approaches.} \vspace{5pt}
\label{tab4}
\begin{tabular}{c|ccc|ccc} \hline
             & \multicolumn{3}{c}{$r_{n}$ (fm)} & \multicolumn{3}{c}{$r_{p}$ (fm)} \\ \cline{2-7}
             &     ETF   &   EVM   &  HF    &   ETF  &   EVM  &   HF  \\ \hline
   $^{16}$O  &    2.54   &  2.57   & 2.67   &  2.51  &  2.65  &  2.70 \\
   $^{40}$Ca &    3.28   &  3.29   & 3.38   &  3.28  &  3.38  &  3.43 \\
   $^{48}$Ca &    3.54   &  3.61   & 3.60   &  3.40  &  3.45  &  3.45 \\
   $^{56}$Ni &    3.62   &  3.67   & 3.64   &  3.64  &  3.77  &  3.70 \\
   $^{78}$Ni &    4.14   &  4.21   & 4.18   &  3.91  &  3.97  &  3.90 \\
   $^{90}$Zr &    4.24   &  4.28   & 4.28   &  4.18  &  4.21  &  4.22 \\
  $^{100}$Sn &    4.33   &  4.36   & 4.34   &  4.38  &  4.48  &  4.42 \\
  $^{132}$Sn &    4.84   &  4.87   & 4.86   &  4.66  &  4.71  &  4.65 \\
  $^{208}$Pb &    5.56   &  5.58   & 5.58   &  5.44  &  5.48  &  5.43 \\ \hline
\end{tabular}
\end{center}
\end{table}

The densities, mean fields and effective masses for neutrons and protons are the outcome of
a restricted variational calculation of the semiclassical HF energy at ETF
level. These energies computed for a set of doubly-magic nuclei are reported in Table \ref{tab2}.
The HF energies, computed quantally as mentioned in Section~II  and explained 
in detail in \cite{soubbotin03}, are also given in the same Table. From the comparison between
the HF and ETF results, we can see that for some nuclei the ETF energy over-binds the HF calculation.
The semiclassical values go through the average of oscillating quantal energies. Therefore, it is natural that
ETF  sometimes gives higher, sometimes lower values than the quantal results. 

According to the Strutinsky Energy theorem \cite{ring-schuck,brack97} the HF energy splits in a large 
smoothly varying part plus a small oscillating shell correction that strongly depends on the considered nucleus.
In our semiclassical calculation the smooth part of the energy is provided by the ETF value. A very efficient
way to recover shell corrections is the so-called expectation value method (EVM) introduced by Bohigas and
collaborators more than forty years ago \cite{bohigas76}. This method consists of including the
shell corrections perturbatively on top of the semiclassical calculation by performing a single HF run using
the semiclassical mean fields computed with the semiclassical densities. As stated in \cite{bohigas76}, the 
EVM treats basically the difference between the HF and ETF kinetic energies as a perturbation and allows to 
recover the shell effects missed in the semiclassical calculation. 
The EVM results obtained starting from our variational ETF calculation are also reported in Table \ref{tab2}. 
We can see that the EVM reproduces the full HF calculations accurately, the differences being always smaller 
than 1\% in all the considered nuclei. The EVM can also reproduce the HF densities in a rather precise way
as it can be seen in Fig. \ref{dens}, where in addition to the ETF and HF densities, we also plot the
densities generated by the EVM. The {\it rms} radii of densities of a set of spherical nuclei, along 
with those included in Fig. \ref{dens}, are presented in Table \ref{tab4}, for all the three approaches
(ETF, EVM and HF). As expected, the radii are quite close to each other in all the three
cases.

\section{Summary and Conclusions}

Extensive ETF calculations for a large number of even-even nuclei spanning the 
entire periodic table have been carried out using the Gogny D1S interaction, with 
the aim to deduce analytic average mean-field potentials from the D1S force with 
space-dependent effective mass, $m^*(r)$. In order to achieve this, the energy density
at ETF level including second-order $\hbar$ corrections is minimized using trial neutron
and proton densities parametrized as generalized Fermi distributions.
The resulting densities are used to compute quantities such as mean-field, spin-orbit potentials
and effective masses for each kind of particles. For each kind of nucleons the mean-field has
 been computed in two different ways. In the first one, the kinetic energy operator includes the 
space-dependent effective mass, whereas in the second one the effective mass contributions from 
the kinetic energy operator have been subsumed in the potential energy. The former approach 
has the advantage that the resulting analytical potentials are directly comparable with the Hartree-Fock
potentials, whereas, the potentials obtained in the latter case, are similar to the phenomenological
potentials available in the literature.
We have demonstrated that 
the mean fields, spin-orbit potentials and position-dependent effective masses
can be parametrized accurately using a simple modified Fermi function. It has
been further shown that the parameters appearing in these Fermi-function forms can be systematized
as functions of neutron and proton numbers up to  very high degree of precision. The resulting
semi-phenomenological potentials and other quantities are found to be very close to those
obtained by the explicit ETF calculation with the D1S force. Our fitted densities, mean field potentials 
and effective masses are ready for use for any kind of nucleus, also the deformed ones albeit by suitable 
modifications of the distance function in the Woods-Saxon form factor. To our knowledge no such potential with effective 
mass can be found in the literature. The fitted neutron and proton densities may be useful in folding model 
calculations to obtain the nucleon-nucleus and nucleus-nucleus optical potentials \cite{bui15}. The knowledge of
these fitted quantities is an advantage for accurate pairing calculations, where it is well known that effective 
masses play an important role \cite{decharge80,kucharek89}. Finally let us mention that these fitted mean-field potentials 
can directly be used to get accurate shell correction energies in Mic-Mac models for nuclear mass tables, as it will be 
shown in the follow-up paper. 
\vskip\baselineskip 
\appendix{{\bf Appendix 1: Additional details on the Extended Thomas-Fermi method with
finite-range forces}}

Using the semiclassical particle, kinetic energy and spin densities (Eqs. (4-6)) given in Section II for 
each type of nucleons, the kinetic, zero-range, Coulomb and spin-orbit contributions to the 
ETF energy density functional can be easily obtained as 
\begin{equation}
{\cal H}_{kin}({\bf R}) = \frac{\hbar^2}{2m} \bigg(\tau_n({\bf R}) + \tau_p({\bf R})\bigg)
\label{eq11}
\end{equation} 
\begin{eqnarray}
 {\cal H}^{nucl}_{z.range}({\bf R}) &=& \frac{t_3}{4}\rho^{\alpha}({\bf R})
\bigg[(2+x_3)\rho^2({\bf R}) \nonumber \\ 
 &-&  (2x_3+1)\bigg(\rho^2_n({\bf R}) + \rho^2_p({\bf R})\bigg)\bigg]
\label{eq12}
\end{eqnarray}

\begin{eqnarray}
 {\cal H}_{Coul}({\bf R}) &=& \frac{1}{2}\rho_p({\bf R})\int d{\bf R'}\,
\frac{\rho_p({\bf R - R'})}{\vert {\bf R - R'}\vert} \nonumber \\ 
&-& 
\frac{3}{4}\left(\frac{3}{\pi}\right)^{1/3}\rho^{4/3}_p({\bf R})
\label{eq13}
\end{eqnarray}

\begin{eqnarray}
{\cal H}_{s.o}({\bf R}) &=& \frac{1}{2}W_0\bigg[\rho({\bf R}){\bf \nabla}\cdot {\bf J}({\bf R})
 \nonumber \\ &+& \rho_n({\bf R}){\bf \nabla}\cdot {\bf J}_n({\bf R})  
+ \rho_p({\bf R}){\bf \nabla}\cdot {\bf J}_p({\bf R})\bigg].
\label{eq14}
\end{eqnarray}

The contributions to the HF energy density due to the finite-range part of the force
are collected in the direct and exchange terms. The direct energy reads: 
\begin{widetext}
\begin{equation}  
{\cal H}^{nucl}_{dir}({\bf R}) = \frac{1}{2} \sum_{i=1}^{i=2}\int d{\bf R'}
v(\vert{\bf R - R'}\vert)\bigg[
X_{d1,i}\bigg(\rho_n({\bf R})\rho_n({\bf R'})
 + \rho_p({\bf R})\rho_p({\bf R'})\bigg) 
+ X_{d2,i}\bigg(\rho_n({\bf R})\rho_p({\bf R'})+\rho_p({\bf R})\rho_n({\bf R'})\bigg)
\bigg],
\label{eq15}
\end{equation}
\end{widetext}
where $X_{d1,i}=W_i+B_i/2-H_i-M_i/2$ and $X_{d2,i}=W_i+B_i/2$, respectively.

The HF exchange energy density is given by
\begin{equation}
{\cal H}_{HF}^{exch} = \frac{1}{2} \int V_ {exch}^{nucl}({\bf R},{\bf s}) 
\rho({\bf R},{\bf s}) \,
d{\bf R} \,d{\bf s} \, ,
\label{eq22}
\end{equation}
where ${\bf R}$ and ${\bf s}$ are the center-of-mass and relative coordinates, respectively,  
$\rho({\bf R},{\bf s})$ the ETF one-body density matrix (Eq. (\ref{eq7})) and 
$V_ {exch}^{nucl}({\bf R},{\bf s})$ the exchange potential, which is defined as 
\begin{equation}
V_ {exch}^{nucl}({\bf R},{\bf s})= - v({\bf R},{\bf s}) \rho({\bf R},{\bf s}),
\label{eq23}
\end{equation}
where $v$({\bf R},{\bf s}) is the finite-range nucleon-nucleon interaction, which in the case of the Gogny 
interaction is given by the first term of Eq.(\ref{eq1}). In the ETF approximation the exchange energy density 
consists of two terms. The first one corresponds to the Thomas-Fermi ($\hbar^0$) contribution,
 which in the Slater approach is given by
\begin{widetext}
\begin{eqnarray}
{\cal H}^{nucl}_{exch,0}({\bf R}) =  \sum_{i=1}^{i=2} \int \,d{\bf s}\,v(s)\bigg\{&\frac{1}{2}&
X_{e1,i}\bigg[\left(\rho_n({\bf R})\frac{3j_1(k_{F_n}({\bf R})s)}{k_{F_n}({\bf R})s}\right)^2
+ \left(\rho_p({\bf R})\frac{3j_1(k_{F_p}({\bf R})s)}{k_{F_p}({\bf R})s}\right)^2\bigg] \nonumber \\
&-& X_{e2,i}\bigg[\rho_n({\bf R})\frac{3j_1(k_{F_n}({\bf R})s)}{k_{F_n}({\bf R})s}
\rho_p({\bf R})\frac{3j_1(k_{F_p}({\bf R})s)}{k_{F_p}({\bf R})s}\bigg]\bigg\}.
\label{eq16}
\end{eqnarray}
\end{widetext}
The second term of the ETF exchange energy density, which contains the $\hbar^2$ corrections, 
is given by 
\begin{widetext}
\begin{eqnarray}
{\cal H}^{nucl}_{exch,2}({\bf R}) = \sum_{q=n,p} \frac{\hbar^2}{2m}\bigg\{\hspace{-12pt} &&(f_q({\bf R})-1)
\left(\tau_q({\bf R})-\frac{3}{5}k_{F_q}^2({\bf R})\rho_q({\bf R}) - 
\frac{1}{4}\Delta\rho_q({\bf R})\right) \nonumber \\
&+& k_{F_q}({\bf R})f'_q({\bf R})\bigg[\frac{1}{27}\frac{({\bf \nabla}\rho_q({\bf R}))^2}
{\rho_q({\bf R})}
- \frac{1}{36}\Delta\rho_q({\bf R})\bigg]\bigg\}.
\label{eq17}
\end{eqnarray}
\end{widetext}

The $\hbar^2$ contribution to the ETF exchange energy contains gradients of the neutron and 
proton densities and is explicitly connected with the nucleon-nucleon interaction through the
 effective mass (Eq. (\ref{eq18})) and its derivative with respect to the position and momentum 
$k$ computed at the Fermi momentum $k_F=(3 \pi^2 \rho({\bf R}))^{1/3}$. Notice that in the 
case of finite-range forces, such as the Gogny interaction, the nucleon effective mass only appears 
explicitly in the energy density when the $\hbar^2$ corrections are taken 
into account. In this case adding to the kinetic energy density (Eq. (\ref{eq11})) the second-order 
exchange energy density one can write:
\begin{widetext}
\begin{eqnarray}
{\cal H}_{kin}({\bf R}) + {\cal H}^{nucl}_{exch,2}({\bf R}) &=& 
\sum_{q=n,p} \frac{\hbar^2}{2m}f_q({\bf R})\tau_q({\bf R})
+ \sum_{q=n,p} \frac{\hbar^2}{2m}\bigg\{(1-f_q({\bf R}))\bigg[\frac{3}{5}k_{F_q}^2({\bf R})
\rho_q({\bf R}) + \frac{1}{4}\Delta\rho_q({\bf R})\bigg]\nonumber \\
&&+k_{F_q}({\bf R})f'_q({\bf R})\bigg[\frac{1}{27}\frac{({\bf \nabla}\rho_q({\bf R}))^2}
{\rho_q({\bf R})} - \frac{1}{36}\Delta\rho_q({\bf R})\bigg]\bigg\}.
\label{eq24}
\end{eqnarray}
\end{widetext}

\vskip\baselineskip 

\appendix{{\bf Appendix 2: Details of Fits}}

As mentioned  in Section 3, here we present the details of parametrizations
developed for the half-density radii, diffusivities, strength parameters,
etc. As discussed there, the densities, mean fields, spin-orbit form factors,
as well as effective masses are assumed to be of appropriately defined 
modified Woods-Saxon form.  

\begin{enumerate}
\item {\bf Densities:} \vskip 0.5\baselineskip 
We take, 
\begin{eqnarray} 
\rho^{(i)} = \rho^{(i)}_{o}\,g\left(r;R^{(i)},a^{(i)}\right)
\end{eqnarray} 
with $i$ = n or p (neutron or proton) and $\rho^{(i)}_{o}$ being determined from the normalization 
condition, and 
\begin{eqnarray} 
g\left(r;R^{(i)},a^{(i)}\right) = \left[f\left(r;R^{(i)},a^{(i)}\right)\right]^{\nu^{(i)}}
\end{eqnarray} 
with
\begin{eqnarray} 
f\left(r;R^{(i)},a^{(i)}\right) = \frac{1}{1 + e^{\left(r - R^{(i)}\right)/a^{(i)}}}
\end{eqnarray} 
The half-density radius and diffusivity parameters for neutrons 
and protons are parametrized as:
\begin{eqnarray}
R^{(n)} &=& 1.2031 \left(1 + 0.0434\,I\right)A^{1/3} - 0.0390 \\
R^{(p)} &=& 1.2559 \left(1 - 0.1200\,I\right)A^{1/3} + 0.1705 \\
a^{(n)} &=& 0.4619 \left(1 + 0.8685 \,I\right) \\
a^{(p)} &=& 0.5656 \left(1 - 0.3175 \,I\right)
\end{eqnarray}
here, $I =\left(N-Z\right)/A$ is the asymmetry parameter. 

\item {\bf Hartree-Fock mean field potentials:} \vskip 0.5\baselineskip 
With the definitions of $f$ and $g$ similar to those defined for the densities,
the mean fields are assumed to be of the form:
\begin{eqnarray} 
V^{(i)}(r) = V_{m}^{(i)}\left[f\left(r;R_{m}^{(i)},a_{m}^{(i)}\right)\right]^{\nu_{m}^{(i)}}
\end{eqnarray} 
where, the function $f$ being similar to the one defined for density distributions. The 
strength, half-density radius and diffusivity of neutrons and protons are 
parametrized as:
\begin{eqnarray}
V_{m}^{(n)} &=& -71.5435\left(1 - 0.3695\,I\right) + 0.7552\,A^{1/3} \\
V_{m}^{(p)} &=& -68.1452\left(1 + 0.4571\,I\right) - 0.2045\,A^{1/3} \\
R_{m}^{(n)} &=&   1.2251\left(1 - 0.1061\,I\right)A^{1/3} + 0.4424 \\
R_{m}^{(p)} &=&   1.1899\left(1 + 0.1359\,I\right)A^{1/3} + 1.2094 \\
a_{m}^{(n)} &=&   0.5587\left(1 + 0.4344\,I\right) \\
a_{m}^{(p)} &=&   0.7636\left(1 + 0.3639\,I\right)
\end{eqnarray}
\item {\bf Phenomenological mean field potentials:} \vskip 0.5\baselineskip 
With a definition similar to the Hartree-Fock mean field potentials, we have:
\begin{eqnarray} 
V_{m}^{(n)} &=& -55.0321 \left(1 - 0.4682\,I\right) + 0.5261\,A^{1/3}\\
V_{m}^{(p)} &=& -51.4006 \left(1 + 0.6258\,I\right) - 0.4422\,A^{1/3}\\
R_{m}^{(n)} &=&   1.2310 \left(1 - 0.1042\,I\right)A^{1/3} + 0.4586 \\
R_{m}^{(p)} &=&   1.1801 \left(1 + 0.1561\,I\right)A^{1/3} + 1.2245 \\
a_{m}^{(n)} &=&   0.4876 \left(1 + 0.7583\,I\right) \\
a_{m}^{(p)} &=&   0.6776 \left(1 + 0.2940\,I\right)
\end{eqnarray} 
where, the symbols have their usual meanings.

\item {\bf Spin-Orbit potential:} \vskip 0.5\baselineskip 
The spin-orbit potential is assumed to be proportional to the derivative
of modified Woods-Saxon form factor. Specifically, we take:
\begin{eqnarray} 
V^{(i)}_{so}\left(r\right) = U_{so}^{(i)}\frac{d}{dr} \left\{
\left[f\left(r;R_{so}^{(i)},a_{so}^{(i)}\right)\right]^{\nu_{so}^{(i)}}\right\}
\end{eqnarray} 
with the various quantities defined in a manner similar to those in 
the density distributions and mean fields. The strength, half-density 
radius and diffusivity for neutrons are protons are parametrized as:
\begin{eqnarray} 
U_{so}^{(n)} &=& 13.1889 \left(1 + 0.3072\,I\right) - 0.1928\, A^{1/3}\\
U_{so}^{(p)} &=& 13.0247 \left(1 - 0.1840\,I\right) - 0.2657\, A^{1/3}\\
R_{so}^{(n)} &=&  1.2201 \left(1 - 0.0014\,I\right)A^{1/3} - 0.0752 \\
R_{so}^{(p)} &=&  1.2360 \left(1 - 0.0452\,I\right)A^{1/3} + 0.1840 \\
a_{so}^{(n)} &=&  0.4836 \left(1 + 0.4937\,I\right) \\
a_{so}^{(p)} &=&  0.5336 \left(1 + 0.0966\,I\right)
\end{eqnarray} 
where, the symbols have their usual meanings.

\item {\bf Effective Mass:} \vskip 0.5\baselineskip 
Finally, the $r$-dependent effective mass is parametrized through:
\begin{eqnarray} 
\frac{m}{m^{*}(r)} - 1 = M^{(i)}_{o}
\left[f\left(r;R_{em}^{(i)},a_{em}^{(i)}\right)\right]^{\nu_{em}^{(i)}}
\end{eqnarray} 
here, $m$ is the average nucleon mass, and the other symbols have their 
usual meanings. The strength, half-density radius and diffusivity for
neutrons and protons are parametrized through:
\begin{eqnarray} 
M_{o}^{(n)} &=&  0.4311 \left(1 - 0.6033\,I\right) \\
M_{o}^{(p)} &=&  0.4349 \left(1 + 0.5772\,I\right) \\
R_{em}^{(n)} &=& 1.2411 \left(1 - 0.0827\,I\right)A^{1/3} + 0.3806 \\
R_{em}^{(p)} &=& 1.1824 \left(1 + 0.1636\,I\right)A^{1/3} + 0.9225 \\
a_{em}^{(n)} &=& 0.3788 \left(1 + 0.4480\,I\right) \\
a_{em}^{(p)} &=& 0.4976 \left(1 + 0.7033\,I\right)
\end{eqnarray} 
here, the symbols have their usual meanings.
\end{enumerate}

The numerical values of the parameters here have been obtained through a
$\chi^2$ minimization process, as discussed in Section 3 of this article.

For the sake of completeness, we summarize the indices ($\nu$) used
for different quantities in Table (\ref{tab3}).

\begin{table}[htb]
\begin{center}
\caption{The indices $\nu$ appearing in the present parametrizations.}\vspace{5pt}
\label{tab3}
\begin{tabular}{ c c c } \hline
 Index             & Quantity where it appears     & Value \\ \hline
$\nu^{(n)}$        &   Neutron density             &  1.5 \\
$\nu^{(p)}$        &   Proton density              &  2.5 \\ \hline
$\nu_{m}^{(n)}$    &   Neutron mean field          &  1.5 \\
$\nu_{m}^{(p)}$    &   Proton mean field           &  3.0 \\\hline
$\nu_{so}^{(n)}$   &   Neutron spin-orbit potential&  1.5 \\
$\nu_{em}^{(p)}$   &   Proton spin-orbit potential &  2.5 \\\hline
$\nu_{em}^{(n)}$   &   Neutron effective mass      &  1.5 \\
$\nu_{so}^{(p)}$   &   Proton effective mass       &  2.5 \\ \hline
\end{tabular}
\end{center}
\end{table}

\begin{acknowledgments}
AB is thankful to Departament de F\'isica Qu\`antica i Astrof\'isica and Institut 
de Ci\`encies del Cosmos, Facultat de F\'isica, Universitat de Barcelona for their kind hospitality.
M.C. and X.V. were partially supported by Grant FIS2017-87534-P from MINECO and FEDER, and by
Grant CEX2019-000918-M from the State Agency for Research of the Spanish Ministry of Science and Innovation 
through the Unit of Excellence Mar\'{\i}a de Maeztu 2020-2023 award to ICCUB.
\end{acknowledgments}

\end{document}